\documentstyle[epsf,psfig]{mn}
\newcommand{\skaco}[1]{\langle{#1}\rangle}
\newcommand{\bm}[1]{\mbox{\boldmath$#1$}}

\begin{document}
\protect
\onecolumn

\title[Cosmological model differentiation through weak gravitational lensing]
{Cosmological model differentiation \\ through weak gravitational lensing}

\author[Antonio C. C. Guimar\~aes]
{Antonio C. C.\ Guimar\~aes\\
  Department of Physics, Brown University, Providence, RI 02912,
  USA; guimar@het.brown.edu 
}

\date{27-Feb-2002, BROWN-HET-1291, astro-ph/0202507}
\maketitle

\begin{abstract} 
We investigate the potential of weak gravitational lensing maps to
differentiate between distinct cosmological models, considering
cosmic variance due to a limited map extension and the presence of noise. 
We introduce a measure of the differentiation between two models under
a chosen lensing statistics. 
That enables one to determine in which circumstances (map size and noise
level), and for which lensing measures two models can be
differentiated at a desired confidence level.
As an application, we simulate convergence maps for three cosmological
models (SCDM, OCDM, and $\Lambda$CDM), calculate several lensing
analyses for them, and compute the differentiation between the models
under these analyses. 
We use first, second, and higher order statistics, including Minkowski
functionals, which provide a complete morphological characterization of the
lensing maps. 
We examine for each lensing measure used how noise affects its description
of the convergence, and how this affects its ability to differentiate
between cosmological models. 
Our results corroborate to the valuable use of weak gravitational
lensing as a cosmological tool.
\end{abstract}
\begin{keywords}
gravitational lensing --- large-scale structure of universe
\end{keywords}


\section{INTRODUCTION}

The mass inhomogeneities in the Universe leave an imprint in the light
traveling through it: that is gravitational lensing.  
The retrieval of this information can be of singular value to narrow
down cosmological models.

The statistical shape distortion of distant galaxies, or cosmic shear, is 
is one of the aspects of gravitational lensing in the weak regime
(small deflection angles), and has been observed by several groups 
(see Mellier 1999 for a review up to this year, and for more recent
measurements see
Van Waerbeke et al. 2000;
Wittman et al. 2000;
Fischer et al. 2000;
Bacon, Refregier \& Ellis 2000; 
Kaiser, Wilson \& Luppino 2000;
Maoli et al. 2001;
Rhodes, Refregier \& Groth 2001;
Van Waerbeke et al. 2001;
Hoekstra, Yee \& Gladders 2001). 
Concomitantly, theoretical works in the area indicated that such 
measurements can be very revealing about the large-scale structure of the
universe (Kaiser \& Squires 1993; 
Bernardeau, Van Waerbeke \& Mellier 1997;
Jain, Seljak \& White 2000; 
Bartelmann \& Schneider 1999, 2001), 
in addition to the study of individual galaxy clusters. 
The main attractive of gravitational lensing being that it probes 
mass directly, avoiding issues such as mass-light bias.

The perspective of a growing number and quality of weak lensing measurements 
stimulates the question of how far the use of weak gravitational
lensing as an astrophysical and cosmological tool can go, and how much
can be achieved.
That ultimately depends on the extension and quality of the lensing
maps, the amount of information that can be extracted from them, and
the ability of this information to differentiate between theoretical
models. 

The aim of this paper is to investigate the potential of weak
gravitational lensing maps to differentiate between different
cosmological models. 
We incorporate limitations resulting from the presence of noise and
cosmic variance for several lensing measures, including a 
morphological analysis, the Minkowski
functionals (Matsubara \& Jain 2001; Sato et al. 2001).

Theoretical predictions for lensing measures can be obtained by two
approaches.
The first is the analytical, which is based on gravitational lens
theory, and resorts frequently on perturbation theory to calculate
expressions for various statistics in an assumed model.
The second approach is to simulate realizations of lensing maps in a
chosen model, and directly measure the quantities of choice.
Here we review the first, and use the second for an application
example.

We obtain that even for modest field sizes (9 degrees$^2$), and noise
at the level of current surveys, the cluster normalized cold dark
matter models (SCDM, OCDM, and $\Lambda$CDM) can be differentiated
with significant confidence. 
Second order statistics (convergence variance and lensing measures
dependent on it) proved to be the best discriminatory analyses for these
models. 
However, we also obtain that higher order statistics (measures that
are sensitive to non-Gaussian features of lensing maps) can
differentiate between some models in certain circumstances. 
This result is particularly important when aiming to
differentiate cosmological models that have the same power spectrum,
but distinct non-Gaussian properties.
We suggest that Minkowski functionals should be included in the row of
statistics normally used to extract information from lensing maps, because
their morphological characterization of maps contains information that
is independent of measures such as the probability distribution
function of the convergence.

The paper is organized as the following. 
In Section \ref{maps} we briefly review how lensing maps can be
generated and analyzed, and we introduce a measure of the
differentiation between two models through a chosen lensing statistic.
In Section \ref{applex} we compare three cosmological models (SCDM,
OCDM, and $\Lambda$CDM) as saw through weak gravitational lensing.
Our conclusions are in Section \ref{conclusions}.

\section{Lensing Map}
\label{maps}
\subsection{Generation}

The image distortion of background galaxies by weak gravitational
lensing can be expressed by the distortion matrix (Bartelmann \&
Schneider 2001)
\begin{equation}
{\cal A}({\bm \theta}) = \left(  
\begin{array}{cc}
1-\kappa-\gamma_1 & -\gamma_2 \\
-\gamma_2 & 1-\kappa+\gamma_1 \\
\end{array}
\right) \; ,
\end{equation}
where $\kappa$ is called the convergence, and 
$\gamma\equiv \gamma_1+i\gamma_2=|\gamma|e^{2i\varphi}$ 
is the shear.

Matrix $\cal A$ transforms a circular source image in an ellipse 
with semi-axis stretched by a factor $(1-\kappa\pm|\gamma|)^{-1}$
from the original radius, and magnification 
$\mu=(\mbox{det} {\cal A})^{-1}=[(1-\kappa)^2-\gamma^2)]^{-1}$.

The shear can be directly estimated from ellipticity measurements of 
background galaxies, and shear and convergence can be mapped in 
each other through
\begin{equation}
\gamma({\bm \theta}) = \frac{1}{\pi} 
\int{{\cal W}({\bm \theta}-{\bm \theta}^\prime)  \kappa({\bm \theta}^\prime)
d^2 {\bm \theta}^\prime } \; ,
\end{equation}
with
\begin{equation}
{\cal W}({\bm \theta})= \frac{-1}{(\theta_1-i\theta_2)^2} \; ,
\end{equation}
or through the relation between their Fourier transforms
\begin{equation}
\tilde{\kappa}({\bm l}) = 
\frac{l_1^2-l_2^2}{l^2} \tilde{\gamma}_1({\bm l}) + 
\frac{2l_1 l_2}{l^2}\tilde{\gamma}_2({\bm l}) \; .
\label{shear-converg}
\end{equation}

We can compute the components ${\cal A}_{ij}$ of the distortion matrix
by calculating the angular deflection of a photon traveling from a
comoving distance $w$ through a gravitational potential $\Phi$
\begin{equation}
{\cal A}_{ij}({\bm \theta}) = -2 \int_0^w{g(w^\prime,w) \partial_i \partial_j 
\Phi({\bm \theta},w^\prime) dw ^\prime } \; ,
\label{distortion}
\end{equation}
where
\begin{equation}
g(w^\prime,w)= \frac{f_K(w^\prime) f_K(w-w^\prime)}{f_K(w)} \; ,
\end{equation}
and $f_K(w)$ is the curvature-dependent
comoving angular diameter distance.

From equation (\ref{distortion}) and using Poisson equation,
\begin{equation}
\nabla^2\Phi=\frac{3}{2}H_o^2\Omega_m \frac{\delta}{a} \; ,
\end{equation}
where $\delta$ is the density contrast and $a$ is the scale factor, we
can write the convergence as 
\begin{equation}
\kappa({\bm \theta}) = \frac{3H_o^2}{2} \Omega_m 
\int_0^{w}{g(w^\prime,w)
\frac{\delta({\bm \theta},w^\prime)}{a(w^\prime)} 
dw^\prime} \; .
\label{convergence}
\end{equation}
The convergence is a weighted projection of the mass inhomogeneities
between source and observer.

In this work we concentrate on the convergence field 
$\kappa({\bm \theta})$ as our lensing map. 
It can be derived from a real or simulated shear map through equation
(\ref{shear-converg}), or computed directly from a N-body simulation of 
$\delta({\bm \theta},w)$ and the use of equation (\ref{convergence}). 
We choose the second approach for our application example at Section
\ref{applex}.

\subsection{Statistics}
\label{mapanalyis}

Here we review some statistics used to characterize the convergence
field,
the angular power spectrum, second and higher order statistics, the
probability distribution function (PDF), and Minkowski functionals.

The convergence field can be expanded in Fourier modes, and their
amplitude averaged for each wavelength.
That is the direct calculation of the angular power spectrum of the 
convergence from a real or simulated lensing map, 
\begin{equation}
P_\kappa(l)\equiv \skaco{|\tilde{\kappa}(\bm{l})|^2} \; .
\end{equation}
An analytical prediction for 
the convergence angular power spectrum can be obtained from
equation (\ref{convergence}), and be expressed as a weighted integral of
the time-evolving density power spectrum  $P_{\delta}(k,w)$
\begin{equation}
2\pi l^2P_{\kappa}(l)=36\pi^2\Omega_m^2l^2 \int_0^{w_o}
{\frac{g^2(w,w_o)}{f_K^2(w)} a^{-2}(w) 
P_{\delta}\left[k=\frac l{f_K(w)},w \right] dw} \; .
\end{equation}
However, the angular power spectrum does not contain all the
statistical information about the $\kappa$-map. All the information in 
the phases of the complex Fourier modes is lost. 
These phases carry the non-Gaussian features of the map, which are
important signatures of non-linear evolution, and also of models of
structure formation containing non-Gaussian initial conditions (as
opposed to most of the inflationary scenarios). 

The Fourier transform of the angular power spectrum of the convergence
is the angular two-point correlation function 
$C_\kappa(r)=\skaco{\kappa(\bm{\theta}) \kappa(\bm{\theta}+\bm{r})}$. 
The convergence field variance $\sigma^2_\kappa\equiv\skaco{\kappa^2}$
can be seen as the value of the angular two-point correlation function
at the origin, $\sigma^2_\kappa = C_\kappa(0)$,  and expressed as 
(Jain \& Seljak 1997; Bernardeau, Van Waerbeke \& Mellier 1997)
\begin{equation}
\skaco{ \kappa^2 } = 
36\pi^2\Omega_m^2 \int_0^{\infty}{kdk}
\int_0^{w_o}{\frac{g^2(w)}{a^2(w)}
P_{\delta}\left(k,w \right) W_2^2\left[k f_K(w)\theta_s \right] dw} \; ,
\end{equation}
where $W_2=2J_1(x)/x$ is the Fourier transform of the top-hat window
function ($J_1$ is the Bessel function of first order). 
Expressions for higher order powers, $\skaco{\kappa^3}$ and
$\skaco{\kappa^4}$, can be similarly obtained as integrals of the
bispectrum and trispectrum, respectively. 
See Hui (1999), and Munshi \& Jain (2000), or  
Cooray \& Hu (2001) for more elegant expressions. 
Related statistics are the skewness 
$S=\skaco{\kappa^3}/\sigma_\kappa^3$, the kurtosis 
$K=(\skaco{\kappa^4}/\sigma_\kappa^4) - 3$, 
or the more general moments 
$S_N=\skaco{\kappa^N}/\sigma_\kappa^{2(N-1)}$.

The probability distribution function contains in principle more
information than the variance and higher order moments of the
convergence field taken individually, which can be calculated from the
PDF itself,
\begin{equation} 
\skaco{\kappa^N} = \int{F(\kappa) \kappa^N d\kappa} \; .
\end{equation}
The PDF can be expanded around a Gaussian through the Edgeworth
approximation (Jusziewicz el al. 1995)
\begin{equation}
F(\kappa)=\frac{e^{-\kappa^2/2\sigma^2_\kappa}}{\sqrt{2\pi}\,\sigma_\kappa}
\left\{ 1 + \sigma_\kappa \frac {S_3}{3!}  
\mbox{H}_3(\kappa / \sigma_\kappa) +  
\sigma^2_\kappa \left[ \frac {S_4}{4!}
\mbox{H}_4(\kappa / \sigma_\kappa) 
+ \frac {S_3^2}{6!} \mbox{H}_6(\kappa / \sigma_\kappa) \right]
+ ... \right\} \; ,
\end{equation}
where $\mbox{H}_\eta$ is the Hermite polynomial of order $\eta$.

Minkowski functionals contain all the morphological information about
a convex body, and for a Gaussian field they can be calculated exactly
(Winitzki \& Kosowsky 1998).  
That makes the Minkowski functionals of the convergence maps a very
interesting statistic to use (Matsubara \& Jain 2001; Sato et al. 2001).
A threshold value $\nu$ defines excursion
sets in which the value of 
$\nu({\bm \theta})\equiv \kappa({\bm \theta})/\sigma_{\kappa}$ is
larger than the threshold: $v_0(\nu)$ is the fractional area of the
map above the threshold, $v_1(\nu)$ the boundary length (per area),
and $v_2(\nu)$ the Euler characteristic. This last functional is
equivalent to the topological genus of the map, roughly speaking, the
number of disconnected regions minus the number of holes.

Approximate expressions for the Minkowski functionals can be obtained through
perturbation theory about the exact expressions for a Gaussian field
(Matsubara 2000)
\begin{eqnarray}
v_{0}(\nu) &\approx&\frac{1}{2}\mbox{erfc}\left( \frac{\nu}{\sqrt{2}} \right)
+ \frac{1}{6 \sqrt{2 \pi}}e^{-\nu^{2}/2}
\sigma_{\kappa}S_3^{(0)}\mbox{H}_{2}(\nu) \;, 
\label{mink0}\\
v_{1}(\nu) &\approx& \frac{1}{8\sqrt{2}}\frac{\sigma_1}{\sigma_{\kappa}}
\mbox{e}^{-\nu^{2}/2}\left\{1+\sigma_\kappa\left[
\frac{S_3^{(0)}}{6}\mbox{H}_{3}(\nu)+\frac{S_3^{(1)}}{3} \mbox{H}_{1}(\nu)
\right]\right\} \;, 
\label{mink1}\\
v_{2}(\nu) &\approx&\frac{1}{2(2 \pi)^{\frac{3}{2}}}
\frac{\sigma_1^{2}}{\sigma_{\kappa}^{2}}
\mbox{e}^{-\nu^{2}/2}
\left\{ \mbox{H}_{1}(\nu)+\sigma_{\kappa}\left[
\frac{S_3^{(0)}}{6} \mbox{H}_{4}(\nu)
+\frac{2S_3^{(1)}}{3} \mbox{H}_{2}(\nu)
+\frac{S_3^{(2)}}{3} 
\right]\right\} \;,
\label{mink2}
\end{eqnarray}
where $\sigma^2_{1}\equiv\skaco{(\nabla\kappa)^2}$, and  
skewness parameters are defined by
$S_3^{(0)}\equiv\skaco{\kappa^3}/\sigma_{\kappa}^4$ ,  
$S_3^{(1)}\equiv-(3/4)\skaco{\kappa^2(\nabla^2\kappa)}/
(\sigma_{\kappa}^2\sigma_1^2)$ and 
$S_3^{(2)}\equiv-3\skaco{(\nabla\kappa\cdot\nabla\kappa)(\nabla^2\kappa)}/\sigma_1^4$.

\subsection{Differentiation}
\label{mapdiff}

Here we introduce a measure aimed to quantify the difference 
between two models, or a model and an observational result, under a
chosen lensing map analysis (Guimar\~aes 2001). 
This quantity is also aimed to allow the comparison of different
lensing statistics, for instance, being able to point which analysis is
the most appropriate to use when trying to differentiate between
models through weak lensing.
We construct one such quantity in Appendix \ref{diff-constr}. 

The {\it differentiation} $\cal D$ between models $A$ and $B$ under the
lensing analysis $Y$  is defined as
\begin{equation}
{\cal D} [Y] \equiv 1 - e^{-\chi^2/2} \; ,
\label{differentiation}
\end{equation}
where 
\begin{equation}
\chi^2 \equiv \frac{1}{N}\sum_i^N \frac{[\bar{Y}_A(p_i)-\bar{Y}_B(p_i)]^2} 
{\sigma_A^2(p_i)+\sigma_B^2(p_i)} \;.
\label{chisquare}
\end{equation}
$\bar{Y}_A(p_i)$ is the mean value of the lensing measure $Y$ for
model $A$ at $N$ values of the analysis parameter $p$, and 
its variance is $\sigma_A^2(p_i)$. Equivalently for model $B$.
Equation (\ref{chisquare}) is a particular discrete form of the more
general expression (\ref{chisqgen}), when the parameter interval is
divided in $N$ equal-size segments.

According to this definition, the differentiation is quantified on a
scale from 0 to 1. 
For models that are similar under a given analysis ${\cal D}[Y]\approx0$, 
and for very distinct models ${\cal D}[Y]\approx 1$.
If one wants to study a set of models, it is desirable to find
which lensing measures give the largest differentiations, so the
discrimination between models is robust. 
On the other hand, if one has an observational result, the interest is
to find which model gives the lowest differentiation for the chosen
analysis. That is the best fit model.  

If we assume that the differentiation has a Gaussian distribution, then it
follows that its variance is 
\begin{equation}
\Delta^2[{\cal D}] = \chi^2 e^{-\chi^2} \; .
\end{equation}


\begin{figure}
\vspace{1cm}
\caption{Convergence maps for SCDM (left), $\Lambda$CDM (center), and
  OCDM (right). The top maps are pure (no noise added), and the bottom
  maps have noise added. All maps are 3$\times$3 degrees$^2$, and smoothed at
  1 arcmin.}
\label{maps-fig}
\end{figure}

\section{Study of Weak Lensing Maps}
\label{applex}

\subsection{Map construction}

We compare three important cosmological
models, SCDM, OCDM, and $\Lambda$CDM (see table \ref{models-table}). 
We used the Hydra N-body code (Couchman, Thomas \&
Pearce, 1995) to 
simulate the large-scale mass distribution in these
models by evolving 128$^3$ particles from a redshift $z=50$ to
$z=0$.
The particle coordinates at juxtaposed boxes from a redshift
$z=1$ (our assumed source redshift) to $z=0$ were saved, and 
projected into a middle plane in each box (the orientation and origin
of the projections were randomized). 
The resulting set of 2D
density contrast fields (in 1024$^2$ grids) at those positions in
redshift space were used as a discrete approximation for the computation
of the convergence field according to equation
(\ref{convergence}), that is usually refered to as multiple-plane lens
method.  

\begin{table}
\caption{Cosmological models considered. Other parameters: $h=0.7$,
  $\Gamma=\Omega_m h$, $\Omega_b=0$, $L_{box}=128h^{-1}Mpc$,
  $l_{soft}=30h^{-1}kpc$, $N_{part}=128^3$.}
\begin{center}
\begin{tabular}{c c c r@{}l }
\hline 
               & $\Omega_m$  & $\Omega_\Lambda$ & $\sigma_8$  & \\
\hline
SCDM           &  1.0  &  0   &   0.56  \\
OCDM           &  0.3  &  0   &   0.84  \\
$\Lambda$CDM   &  0.3  &  0.7 &   0.99  \\
\hline
\end{tabular}
\end{center}
\label{models-table}
\end{table}

The main source of noise in real lensing maps is the intrinsic
ellipticities of the background galaxies.
We simulated this noise by adding a Gaussian random field to the pure
convergence following Van Waerbeke (2000), which showed that this is a
good approximation.
We then used a top-hat window of radius $\theta_s$ to smooth
the map, which can be described by 
\begin{equation}
\kappa({\bm \theta}) = \kappa_o({\bm \theta}) + n({\bm \theta}) \; ,
\label{noise}
\end{equation}
where $\kappa_0({\bm \theta})$ is the smoothed pure convergence, 
and $n({\bm \theta})$ is the smoothed noise field. 
The noise part is a Gaussian correlated field of
variance $\sigma_n^2$,
\begin{equation}
\sigma_n^2 = \frac{\sigma^2_{\epsilon}}{2n_g\pi\theta_s^2} \; ,
\label{noise-var}
\end{equation}
where $\sigma^2_{\epsilon}=0.16$ was the intrinsic ellipticity
variance adopted, and $n_g=60$ arcmin$^{-2}$ was the mean source galaxy
density assumed. 
The top-hat smoothing of the original Gaussian random field introduces
correlations in the noise at scales bellow twice the smoothing radius.

Figure \ref{maps-fig} shows some realizations of the $\kappa$-maps. 
We generated 25 realizations for each model, however, because we only
used one N-body simulation for each model, the $\kappa$-maps generated
cannot be considered to be totally independent. 
That is not a source of major concern for our objectives in this work.

\subsection{Statistics and differentiation results}

We applied a number of statistical measures to the convergence maps of
the three cosmological models. We considered pure (no noise) and
noisy maps of field sizes 3$\times$3 degrees$^2$, and 1$\times$1
degree$^2$. 
For each analysis we calculated the differentiation $\cal D$ according to 
equation (\ref{differentiation}) between $\Lambda$CDM and SCDM, and
between $\Lambda$CDM and OCDM.

\begin{figure}
\epsfxsize=3.5in
\centerline{\epsfbox{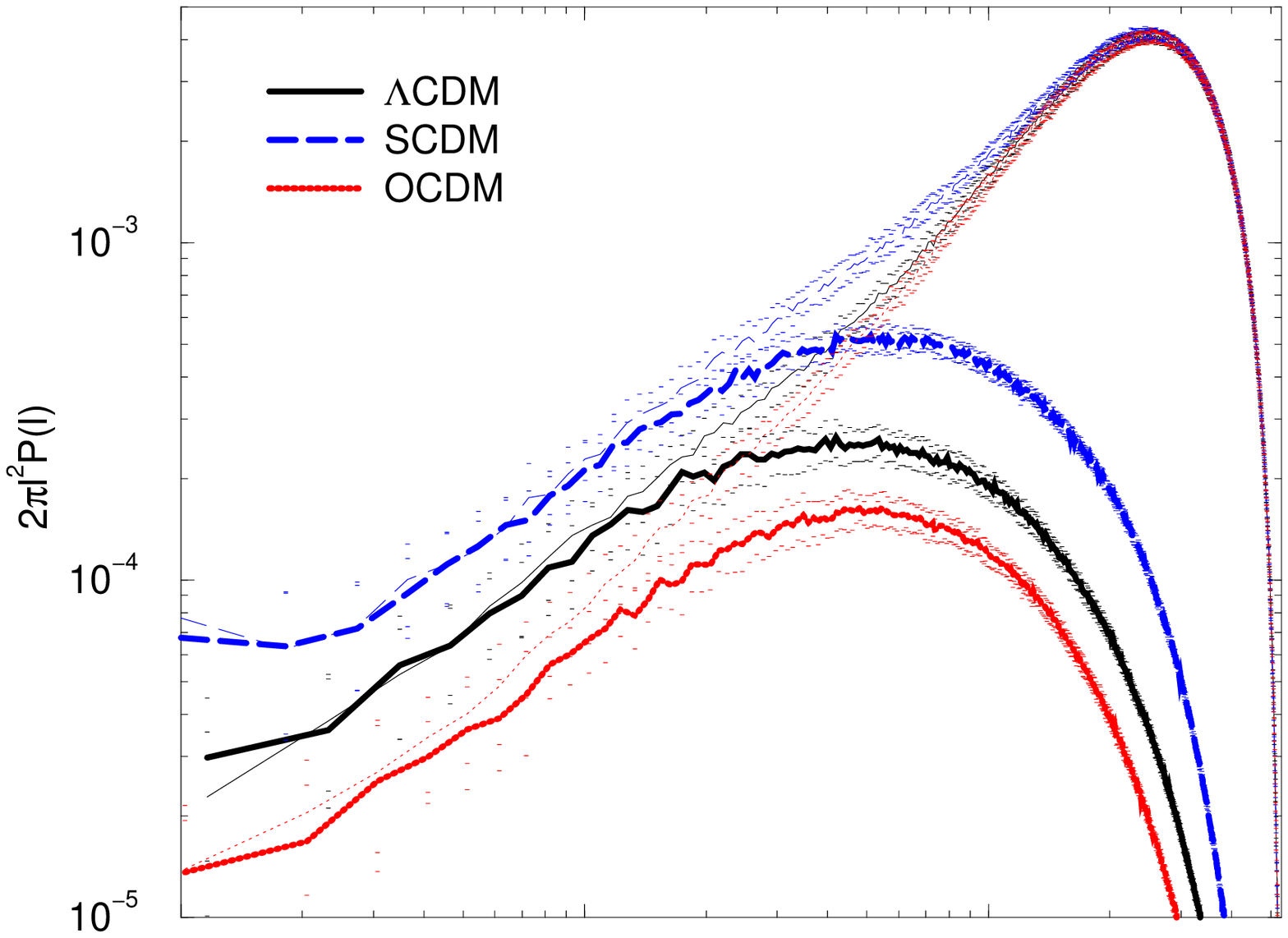}}
\vspace{-1.2cm}
\epsfxsize=3.5in
\centerline{\epsfbox{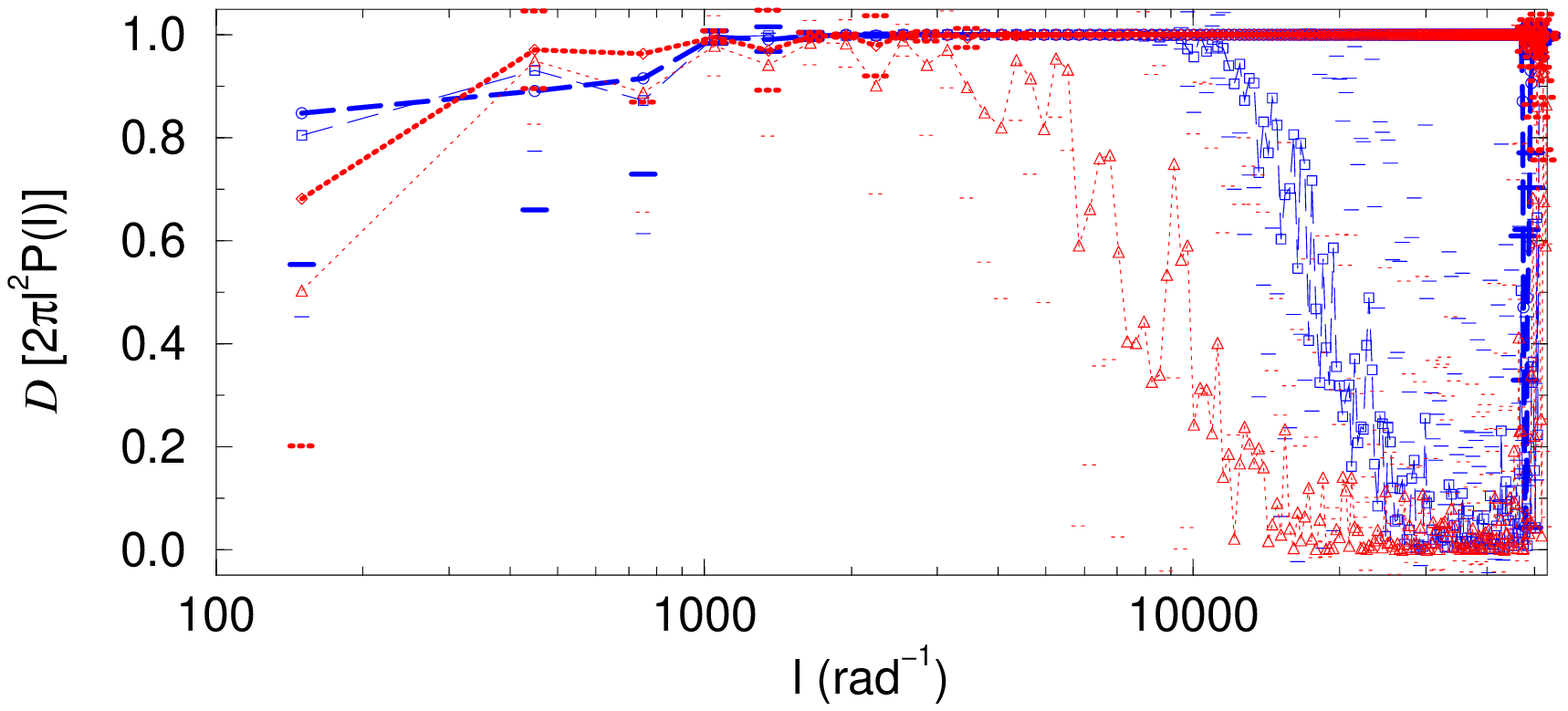}}
\caption{Top panel: angular power spectrum of the convergence field smoothed at
  0.25 arcmin scale. Bottom panel: angular power differentiation
  between models $\Lambda$CDM-SCDM (dashed lines), and
  $\Lambda$CDM-OCDM (dotted lines).
  Thick lines are for pure maps, and thin lines for noisy maps. 
  Only the error-bar tips are show for clarity.}
\label{power-fig}
\end{figure}

Figure \ref{power-fig} shows the results for the angular power
spectrum of the convergence for a field smoothed with $\theta_s=0.25$
arcmin, and its differentiation between models. 
The total power of the convergence field, $P_{\kappa}$, is just given
by the sum of the power of the pure field $P_{\kappa_o}$ and the power
of the noise field $P_n$ (a power law), because the two fields are
uncorrelated (by construction), 
\begin{equation}
P_{\kappa}(l) = P_{\kappa_o}(l) + P_n(l) \; ,
\end{equation}
and its variance can also be fragmented as
\begin{equation}
\Delta^2[P_{\kappa}]=\Delta^2[P_{\kappa_o}]+\Delta^2[P_n] +
\Delta^2[\skaco{\tilde{\kappa}_o \tilde{n}^*}] +
\Delta^2[\skaco{\tilde{\kappa}_o^* \tilde{n}}] \; .
\end{equation}
For low values of the wavenumber $l$ (large scales) noise gives a
small contribution to the power of the convergence, but the variance
of the measured power is large due to the restricted sampling. This
large variance implies a lower differentiation between models at low $l$. 
For high $l$ (small scales) the measured power is suppressed by the
field smoothing at wavenumber values above $l_s \sim
\theta_s^{-1}$. The noise field gives a major contribution to the
total power and power variance, such that the examined models become
indistinguishable as quantified by the low values of ${\cal D}\, [2\pi
l^2 P(l)]$ in this case. 
Note that the analyzed maps have different areas
($\Lambda$CDM 9.6 degrees$^2$, SCDM 15.6 degrees$^2$, and OCDM 12.3
degrees$^2$), therefore the power variances cannot be directly
compared.

\begin{figure}
\leavevmode
\centerline{
\epsfxsize=1.88in
\epsfbox{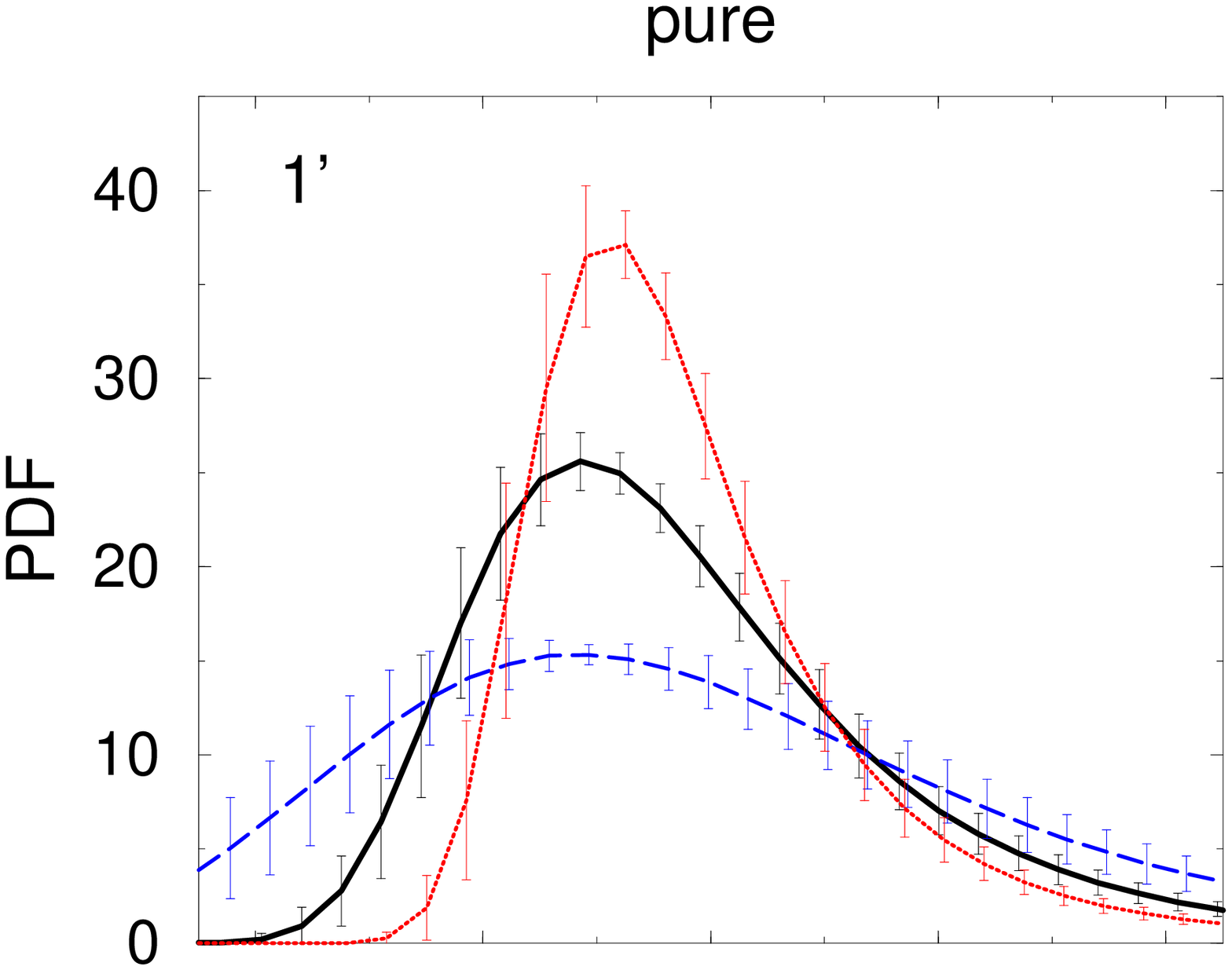}
\hspace{-1.2cm}
\epsfxsize=1.88in
\epsfbox{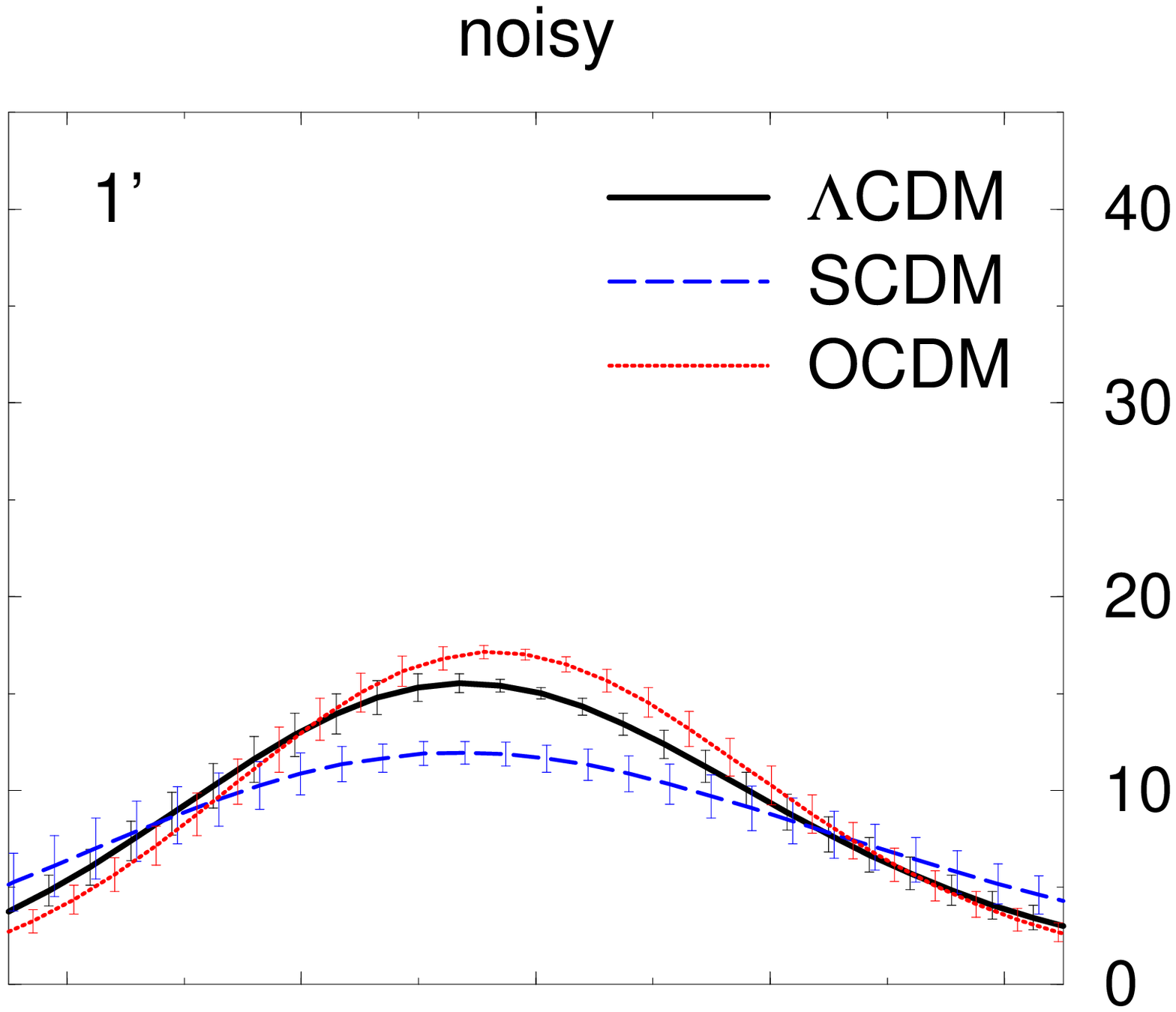}}
\vspace{-1.3cm}
\\
\centerline{
\epsfxsize=1.88in
\epsfbox{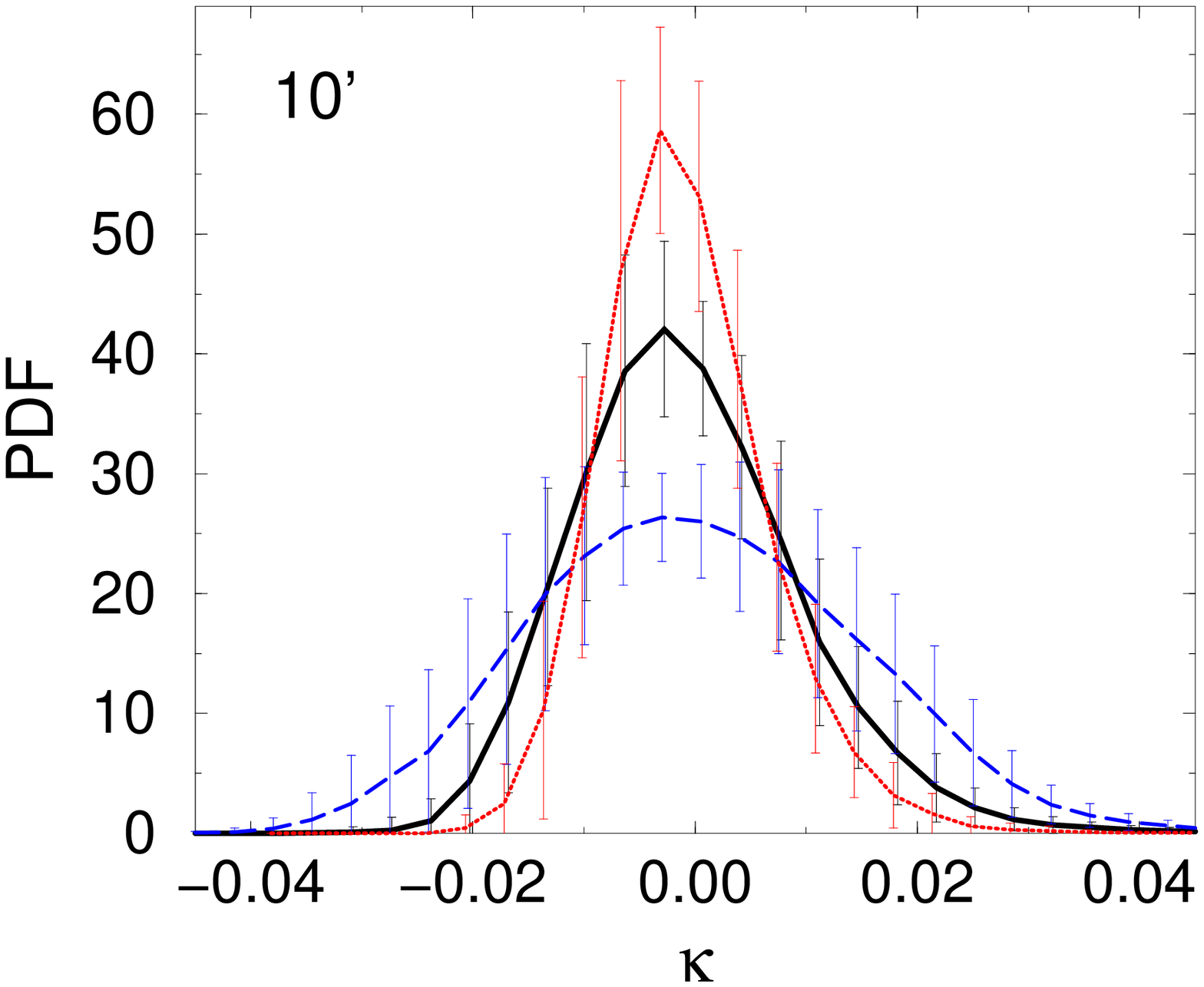}
\hspace{-1.2cm}
\epsfxsize=1.88in
\epsfbox{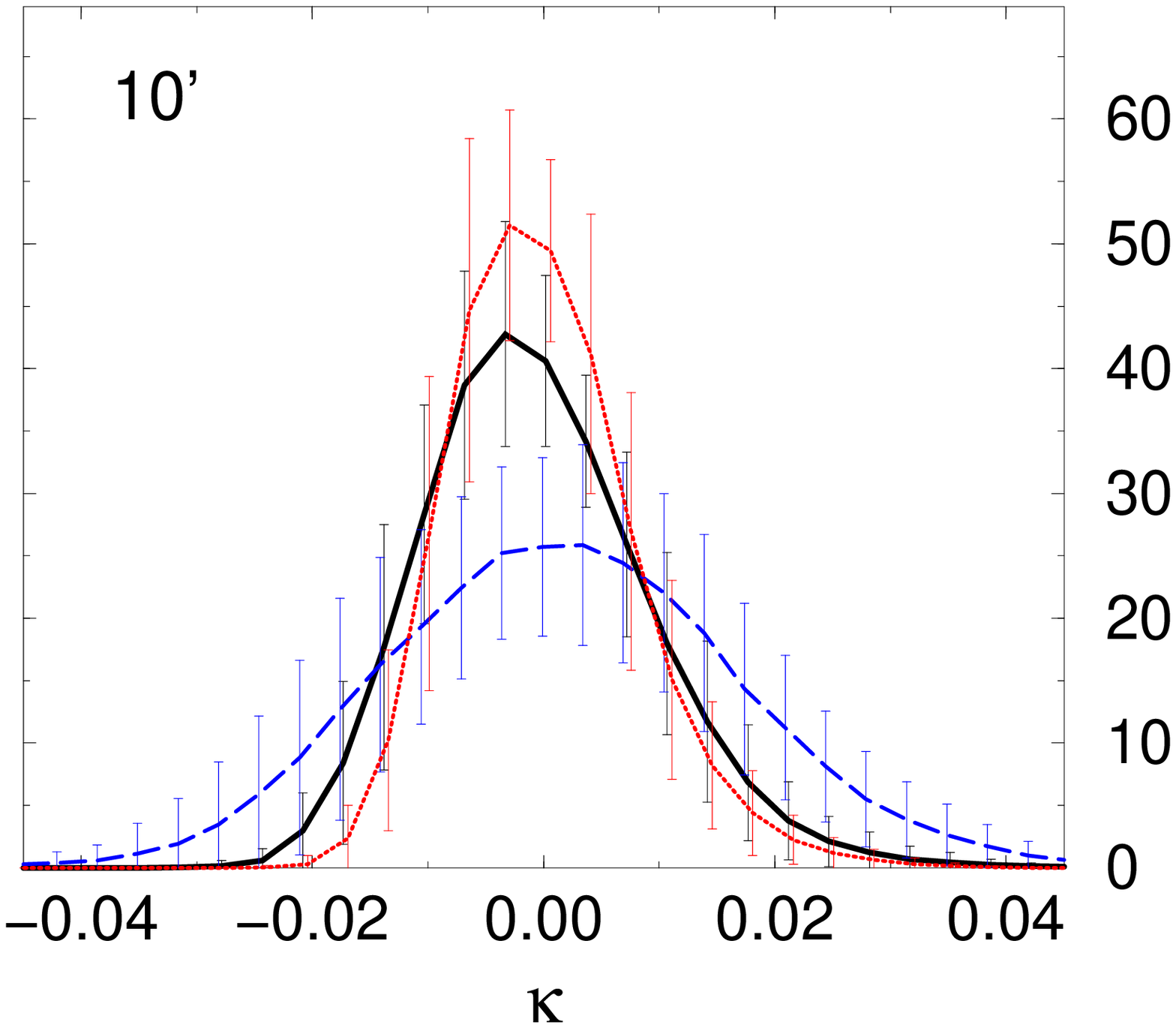}}
\vspace{-0.3cm}
\\
\centerline{
\epsfxsize=1.88in
\epsfbox{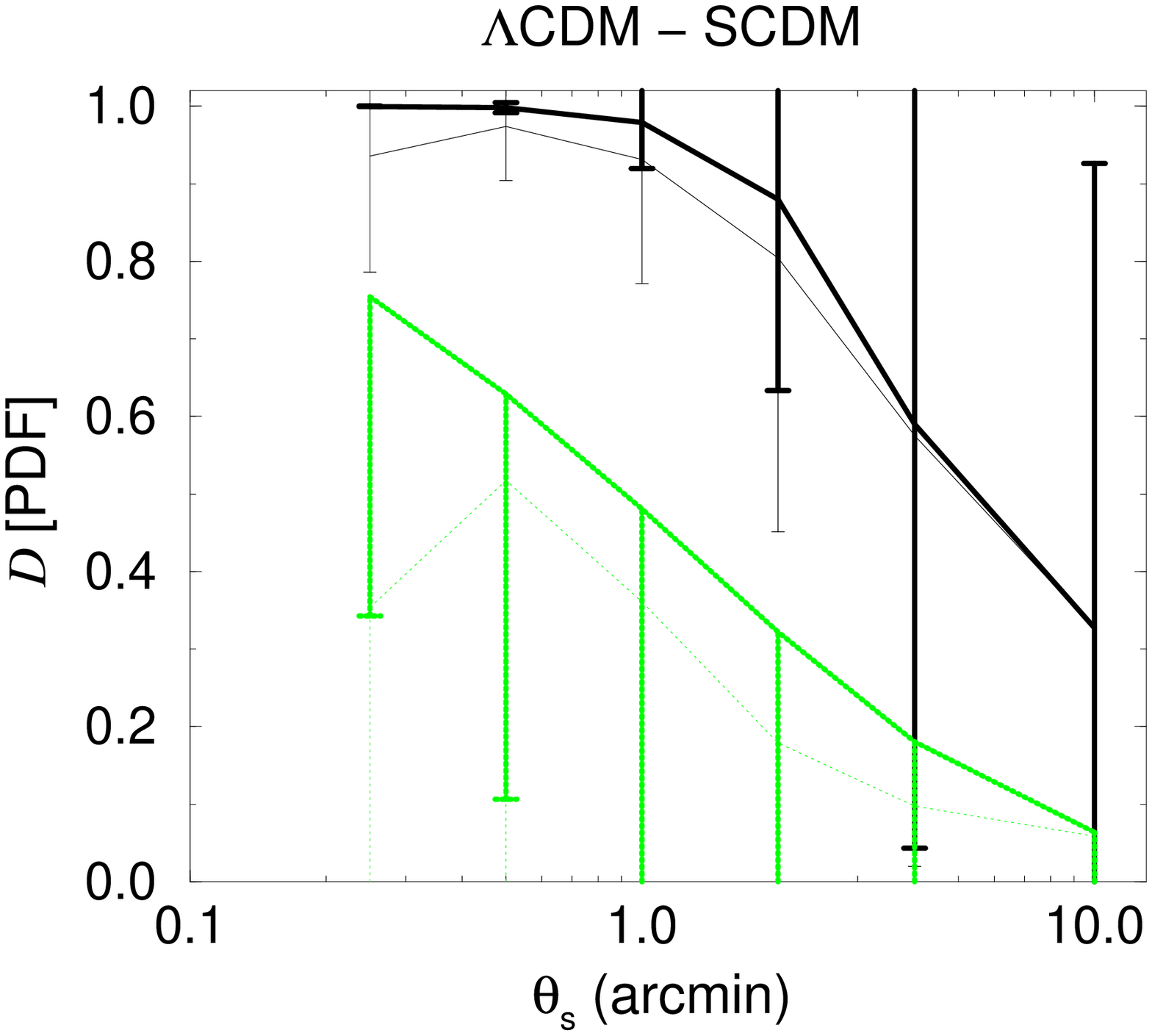}
\hspace{-1.2cm}
\epsfxsize=1.88in
\epsfbox{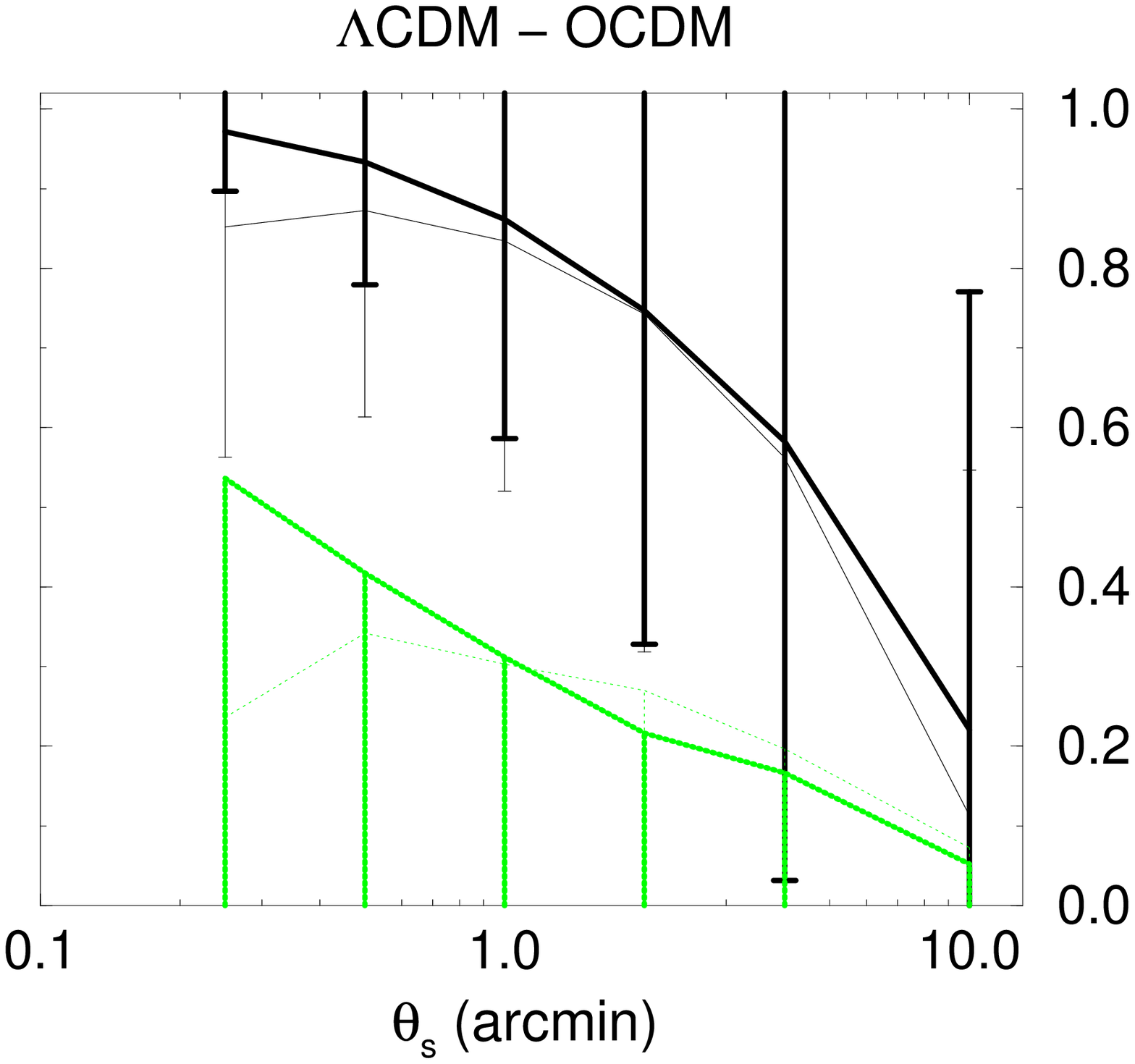}}
\vspace{-0.cm}
\caption{Top panels: convergence probability distribution function for 
  pure (left) and noisy (right) maps at 1 and 10 arcmin smoothing
  scales of 3$\times$3 degrees$^2$ fields.
  Bottom panels: PDF differentiation between
  $\Lambda$CDM-SCDM (bottom left) and  
  $\Lambda$CDM-OCDM (bottom right), as function of the smoothing
  angle $\theta_s$. Thick lines are for pure maps, and thin lines for
  noisy maps; solid lines are for 3$\times$3 degrees$^2$ fields, and
  dotted lines are for 1$\times$1 degree$^2$ fields (the superior error-bars
  for these are not show for clarity).}
\label{PDF-fig}
\end{figure}

\begin{figure}
\centerline{
\epsfxsize=3.5in
\epsfbox{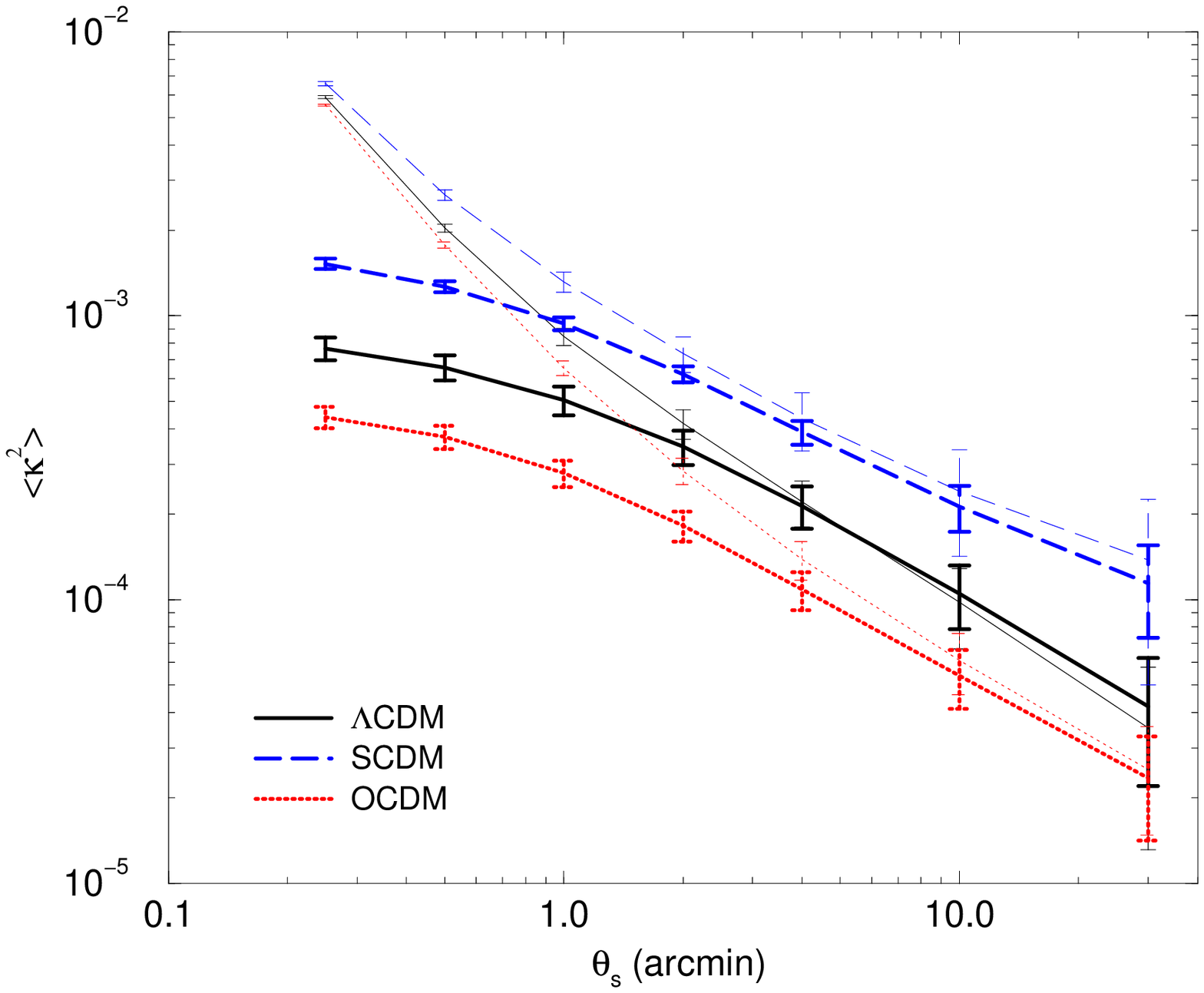}}
\vspace{-0.4cm}
\leavevmode
\centerline{
\epsfxsize=1.88in
\epsfbox{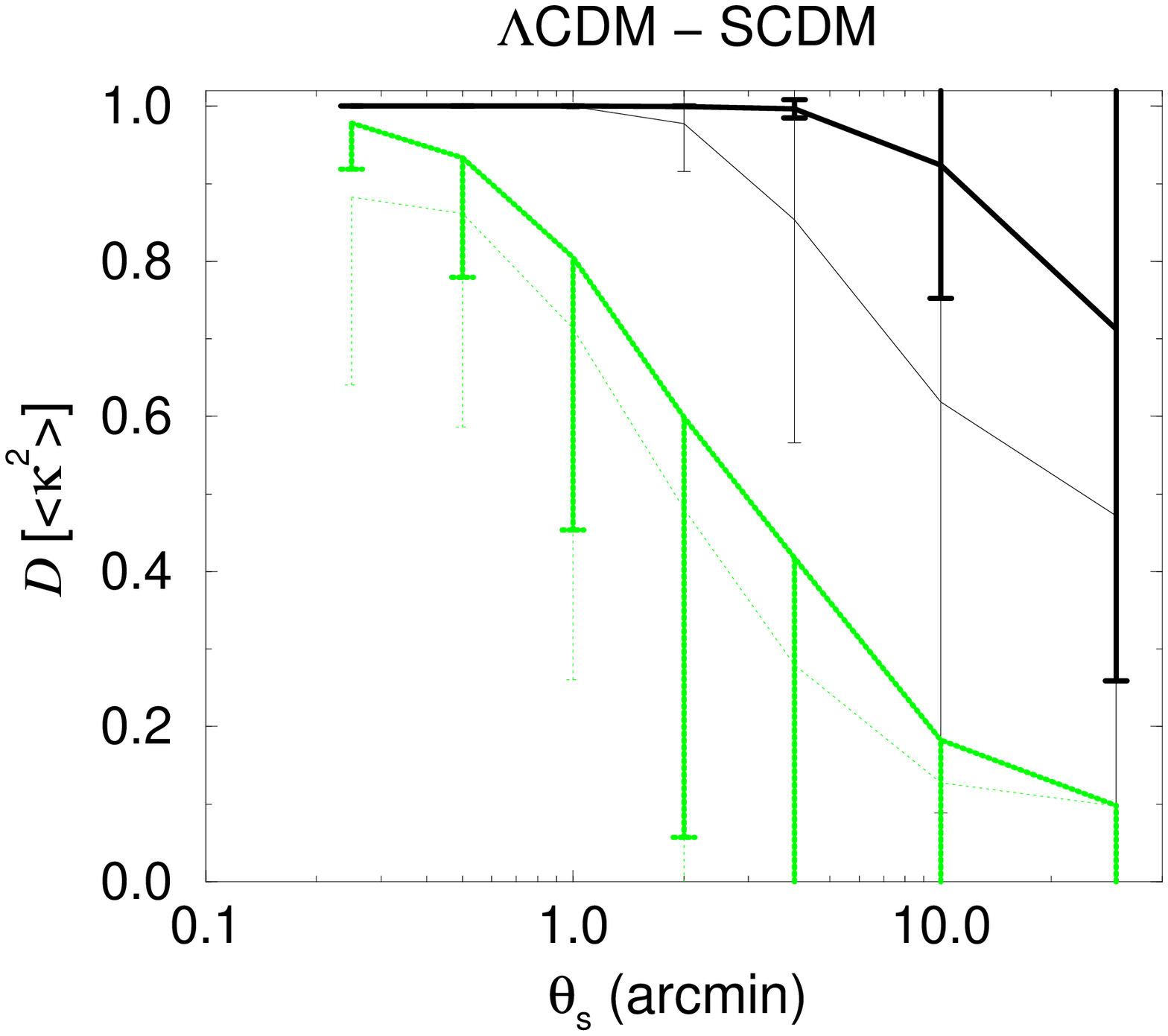}
\hspace{-1.2cm}
\epsfxsize=1.88in
\epsfbox{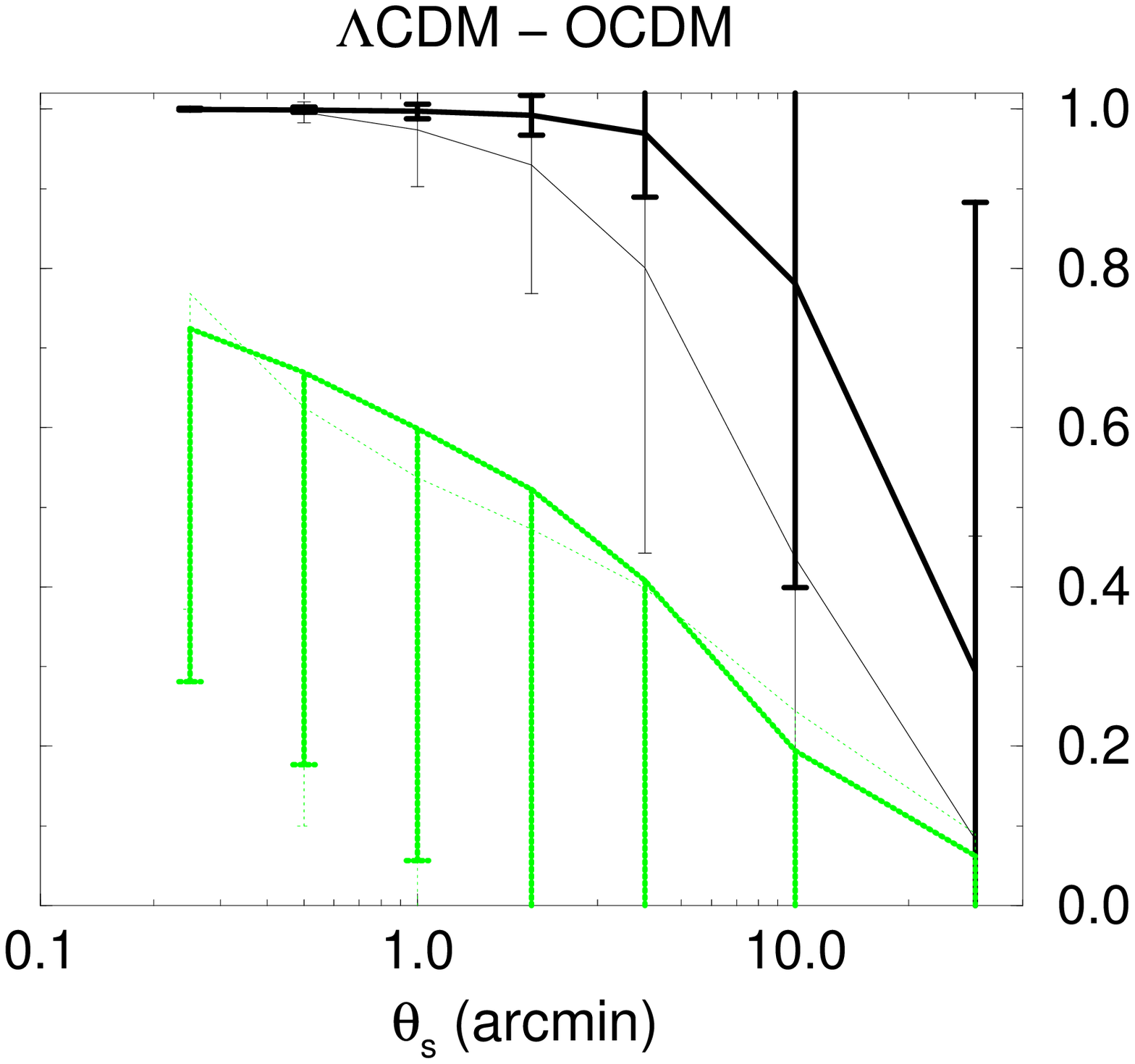}}
\vspace{-0.3cm}
\caption{Convergence variance of 3$\times$3 degrees$^2$ fields (top panel), 
  and convergence variance differentiation between
  $\Lambda$CDM-SCDM (bottom left) and  
  $\Lambda$CDM-OCDM (bottom right), as function of the smoothing
  angle $\theta_s$.
  Thick lines are for pure maps, and thin lines for noisy maps.
  In the bottom panels solid lines are for 3$\times$3 degrees$^2$ fields, and
  dotted lines are for 1$\times$1 degree$^2$ fields (the superior error-bars 
  for these are not show for clarity).}
\label{k2-fig}
\end{figure}

The addition of noise transforms the probability distribution function
for the pure convergence field $F_{\kappa_o}$ 
according to the convolution
\begin{equation}
F_{\kappa}(x)=\int_{-\infty}^{+\infty}{F_{\kappa_o}(y)
\frac{e^{(x-y)^2/\sigma_n^2}}{\sqrt{2\pi}\sigma_n} dy} \; .
\label{pdf-convolution}
\end{equation}
For a small smoothing angle $\theta_s$ the noise variance $\sigma_n^2$
is large, so the convolution has a large effect on the pure
convergence PDF. 
In contrast, for a large $\theta_s$, $\sigma_n^2$ is small, and
$F_{\kappa}$ does not differ substantially from $F_{\kappa_o}$.
This effect is illustrated on Figure \ref{PDF-fig}, which  shows 
the results for the probability
distribution function of the $\kappa$ field, and the integrated
PDF differentiation in the interval $-0.04 < \kappa < 0.04$. 
The differentiation curves indicate that noise reduces the
differentiation between models, but moderately.
Smoothing has a much more visible role in reducing this
differentiation - the convergence maps for different models are
made more homogeneous by smoothing, and therefore made more alike.
The map size also has a major effect on the ability of
the PDF analysis to differentiate between models, because the variance
of the PDF measurement increases with a reduced field size.

Figure \ref{k2-fig} shows the convergence variance 
$\skaco{\kappa^2}$ as a 
function of the smoothing angle, and its model differentiation. 
Because the pure convergence and noise fields are uncorrelated, the
convergence variance can be written as
\begin{equation}
\skaco{\kappa^2 } = \skaco{\kappa^2_o} + \,\sigma_n^2 \; ,
\end{equation}
\begin{equation}
\Delta^2[\sigma^2_\kappa] = \Delta^2 [\sigma^2_{\kappa_o}] + 
\Delta^2[\sigma^2_n] + 2 \Delta^2[\skaco{\kappa_o n} ] \; .
\end{equation}
Noise dominates for small field smoothing, because 
$\sigma_n^2 \propto \theta_s^{-2}$, while the variance of the pure
field decreases with a smaller power 
($\sigma^2_{\kappa_0}\sim \theta_s^{-0.8}$), and is 
negligible for large $\theta_s$ (error-bars are increased though).
The curves for ${\cal D}\,\left[\skaco{ \kappa^2}\right]$
demonstrates that the field variance is a powerful analysis to
discriminate between the models considered, even for small maps.

\begin{figure}
\leavevmode
\epsfxsize=3.4in
\epsfbox{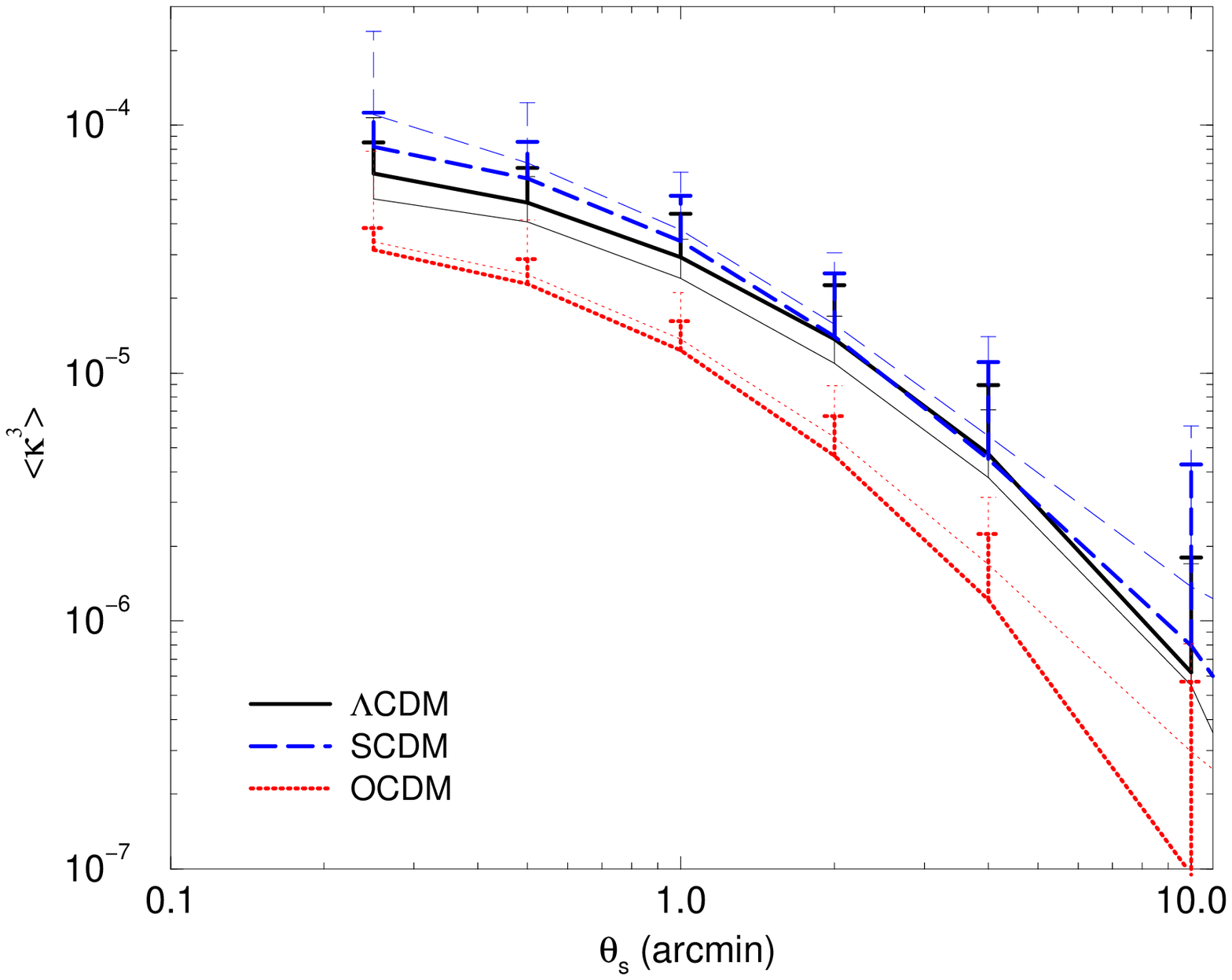}
\epsfxsize=3.4in
\epsfbox{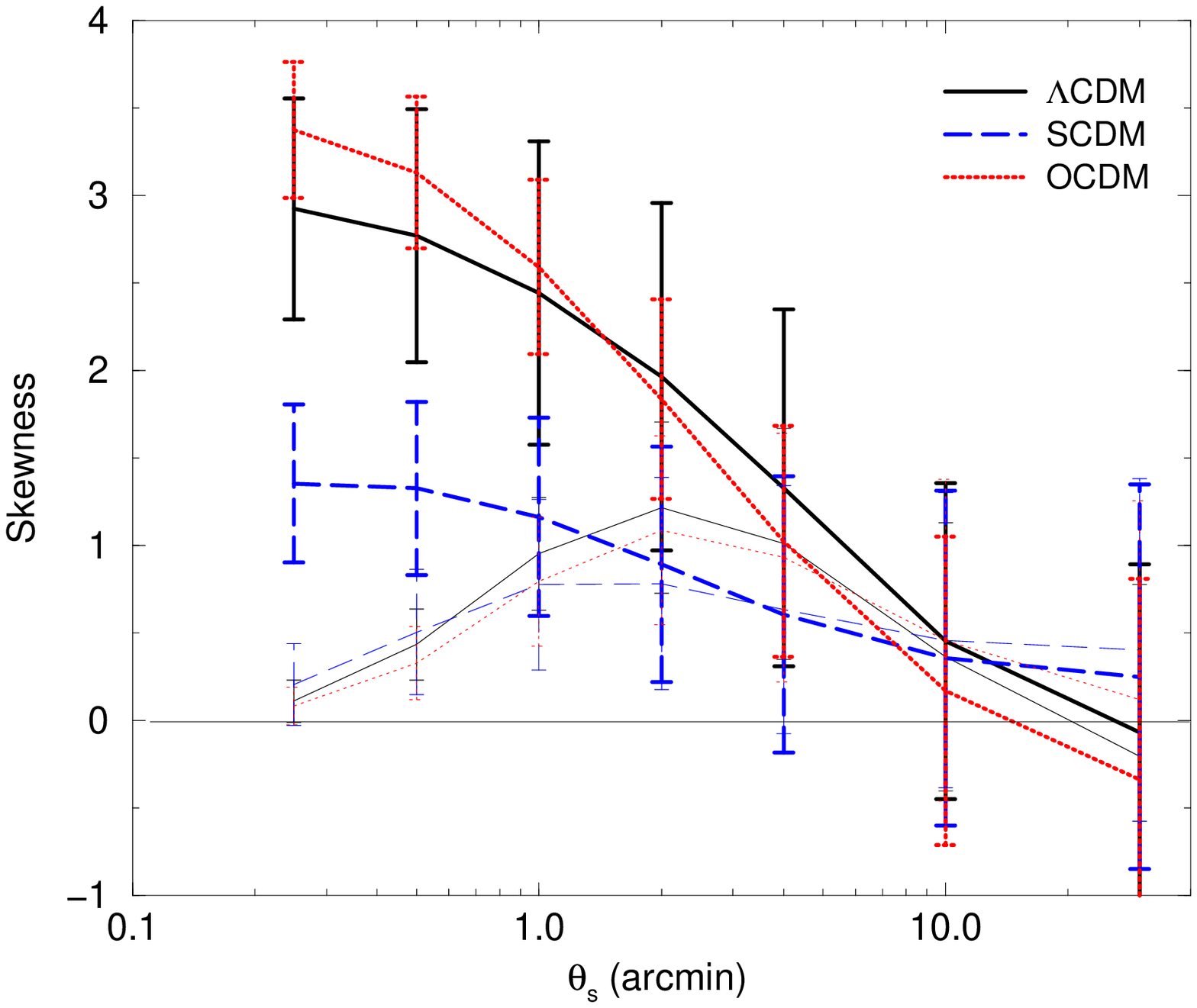}
\vspace{-0.4cm}
\\
\epsfxsize=1.88in
\epsfbox{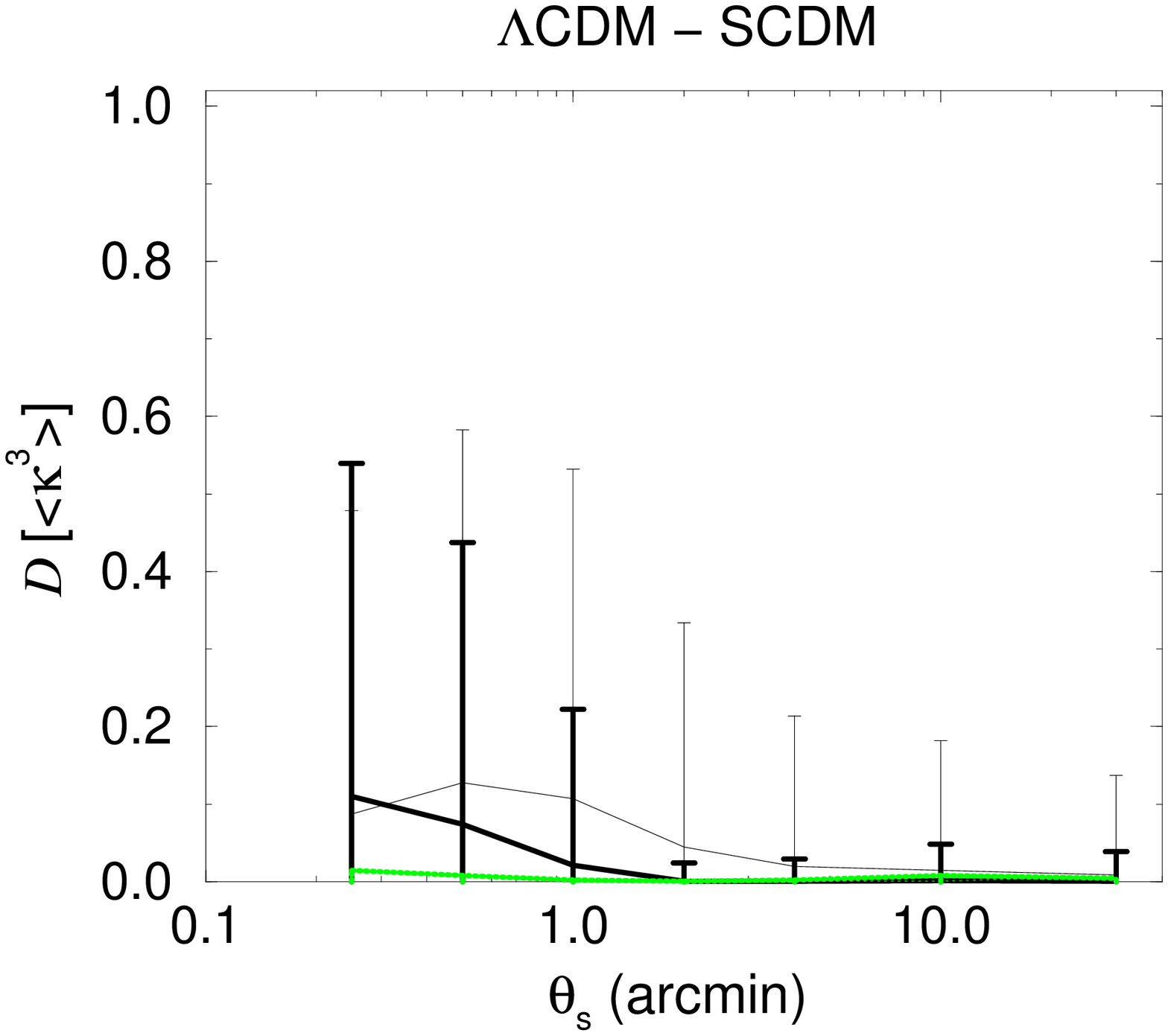}
\hspace{-1.2cm}
\epsfxsize=1.88in
\epsfbox{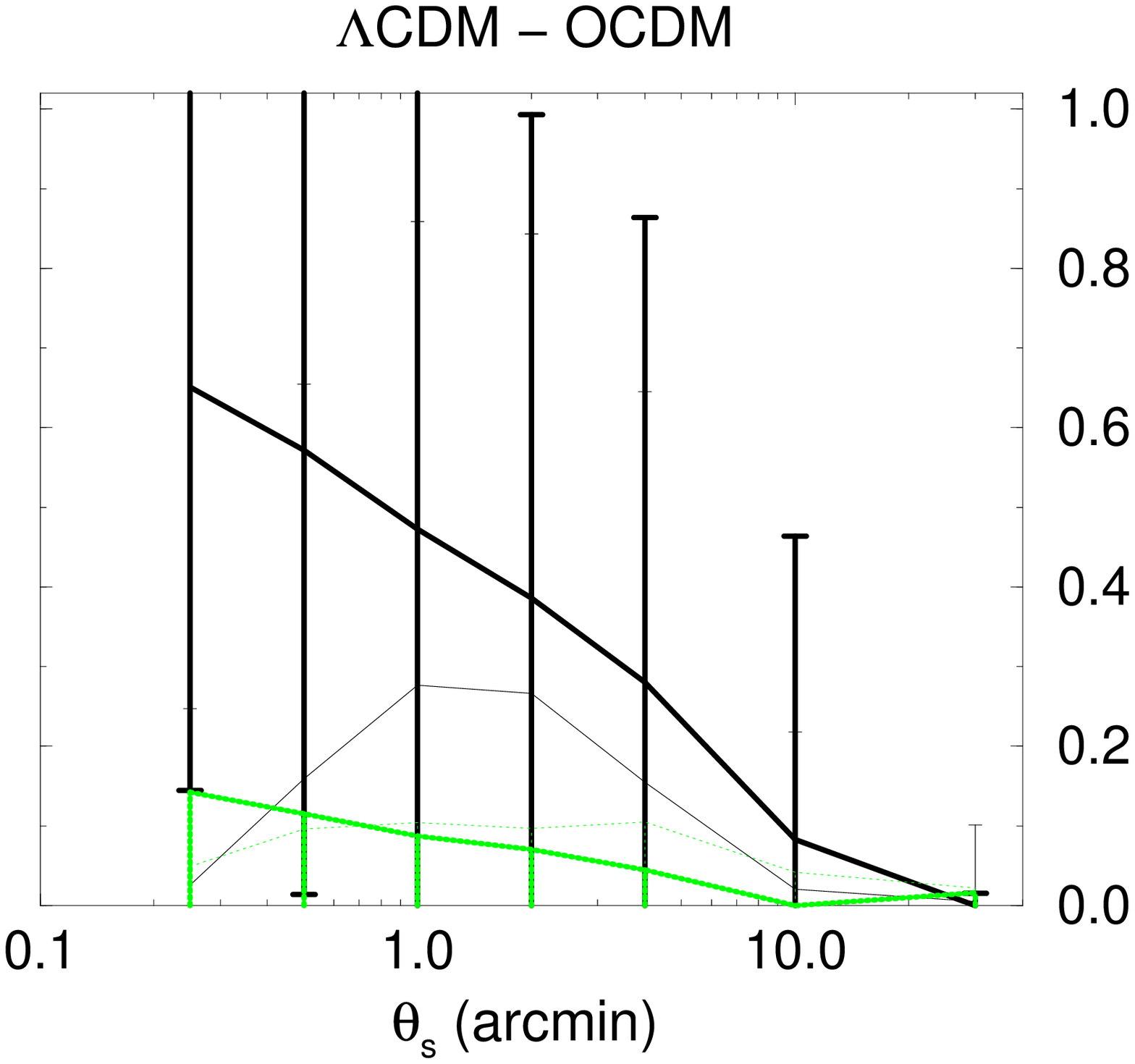}
\hspace{.5cm}
\epsfxsize=1.88in
\epsfbox{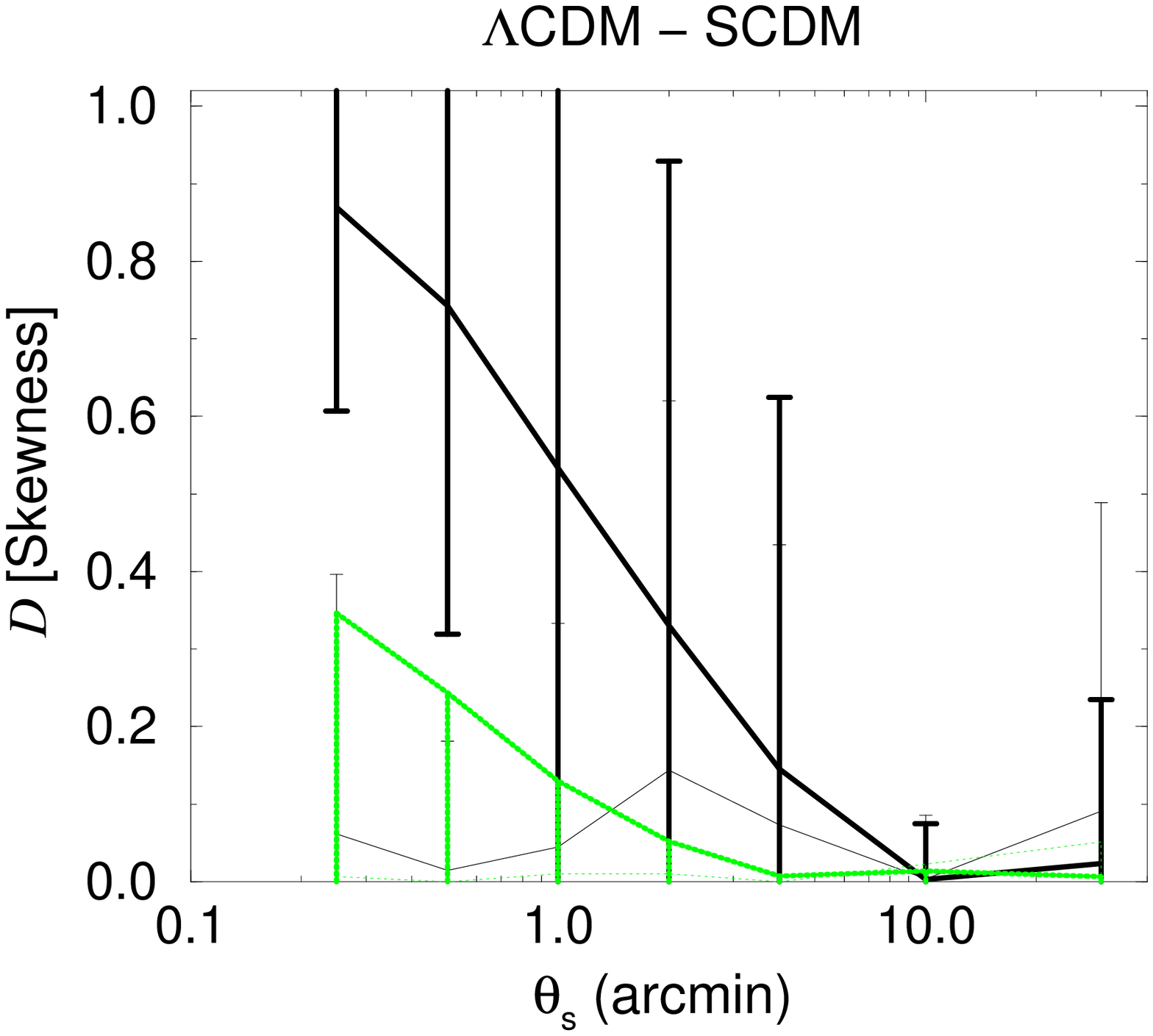}
\hspace{-1.2cm}
\epsfxsize=1.88in
\epsfbox{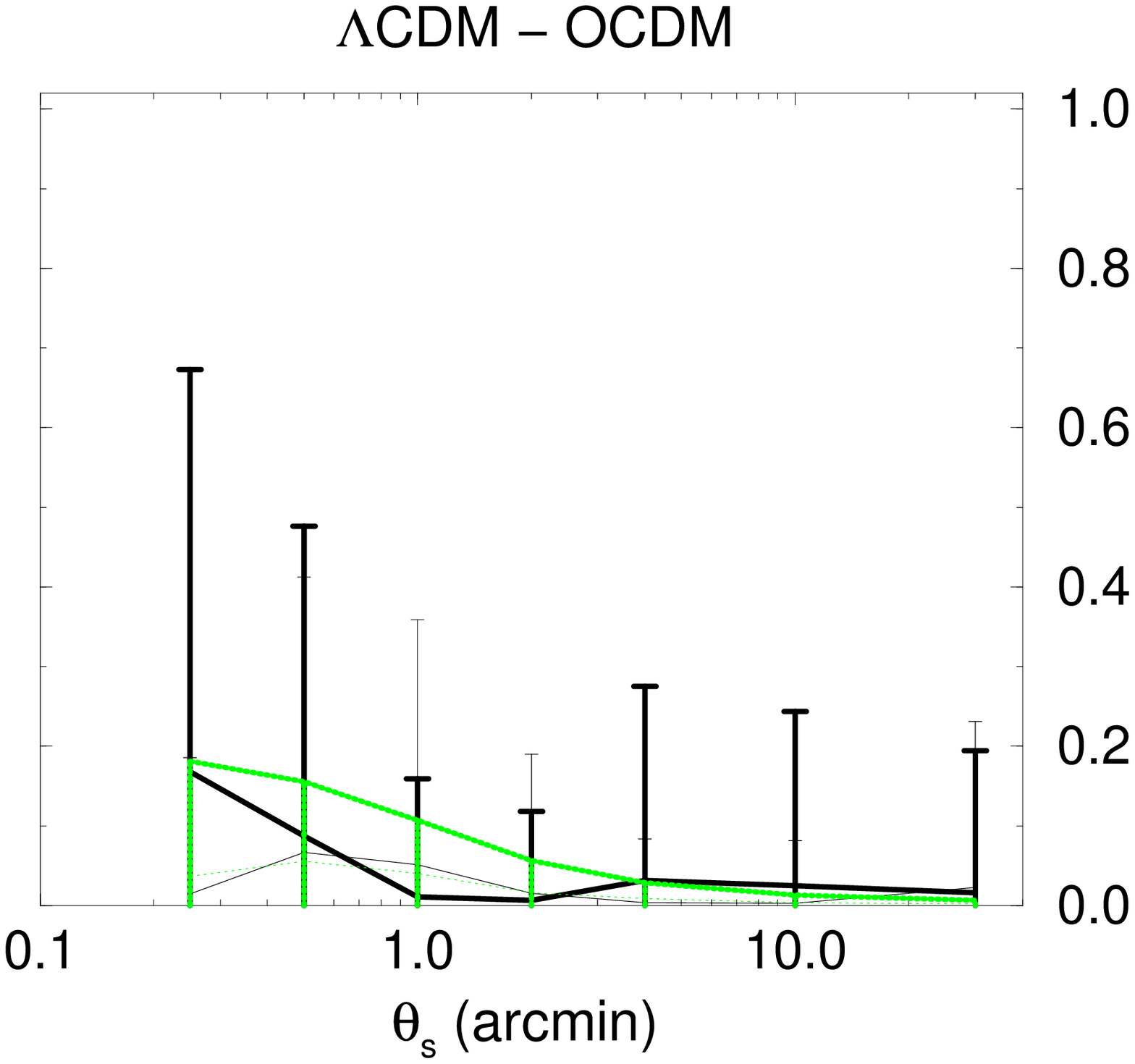}
\vspace{-0.2cm}
\caption{Convergence mean third power (left) and skewness (right). 
  Equal graphic conventions to figure \ref{k2-fig} are used.}
\label{k3-fig}
\end{figure}
\begin{figure}
\leavevmode
\epsfxsize=3.4in
\epsfbox{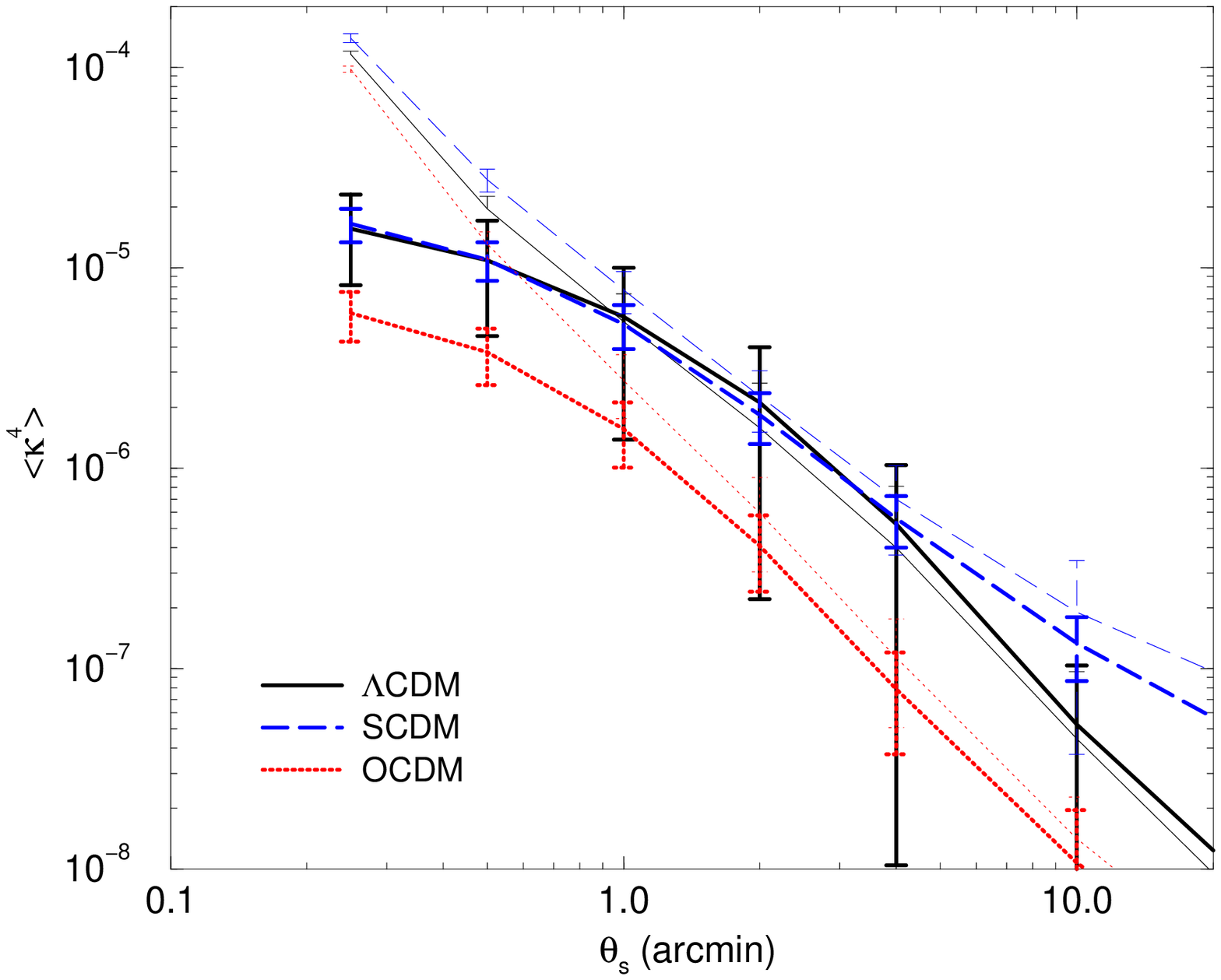}
\epsfxsize=3.4in
\epsfbox{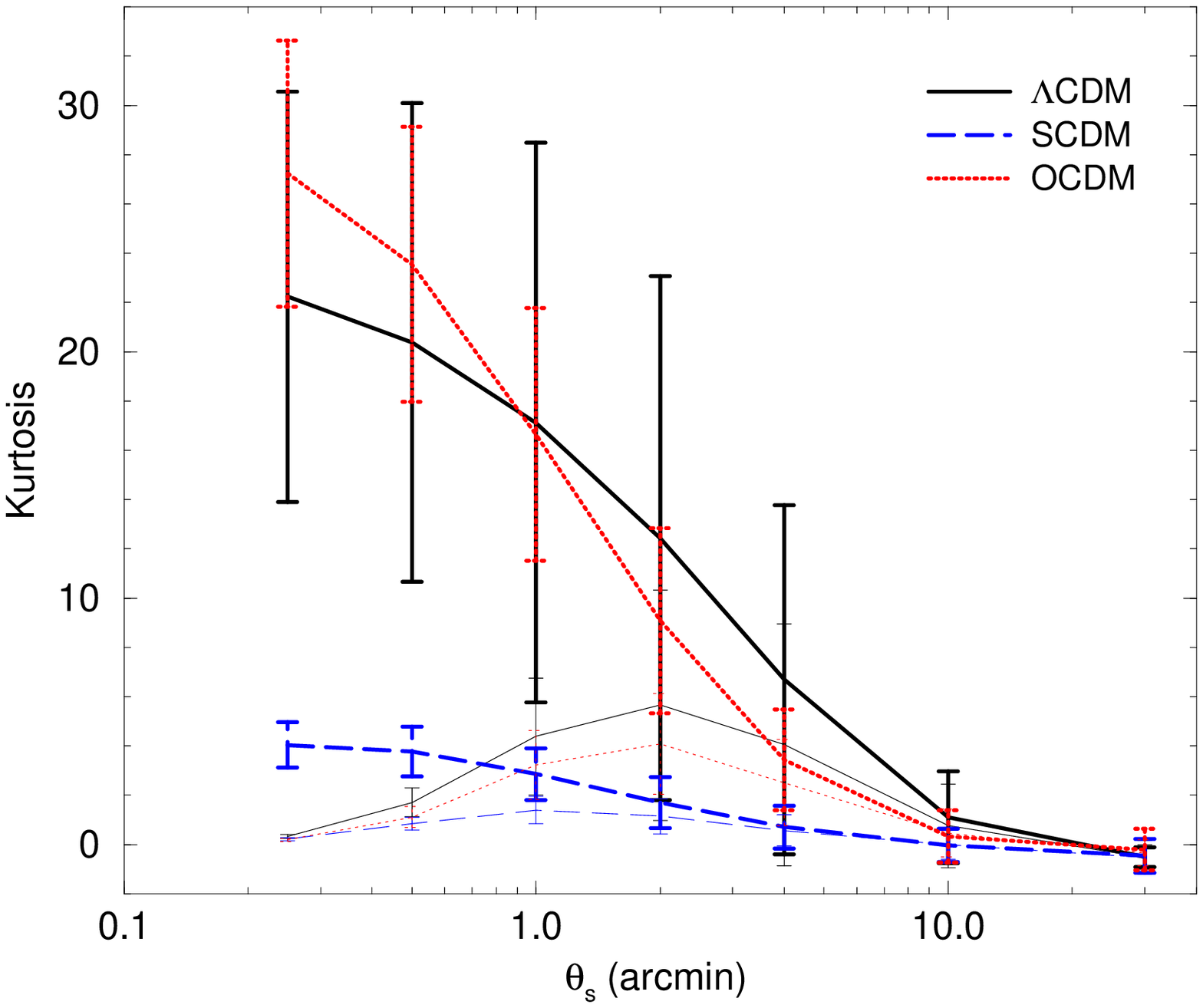}
\vspace{-0.4cm}
\\
\epsfxsize=1.88in
\epsfbox{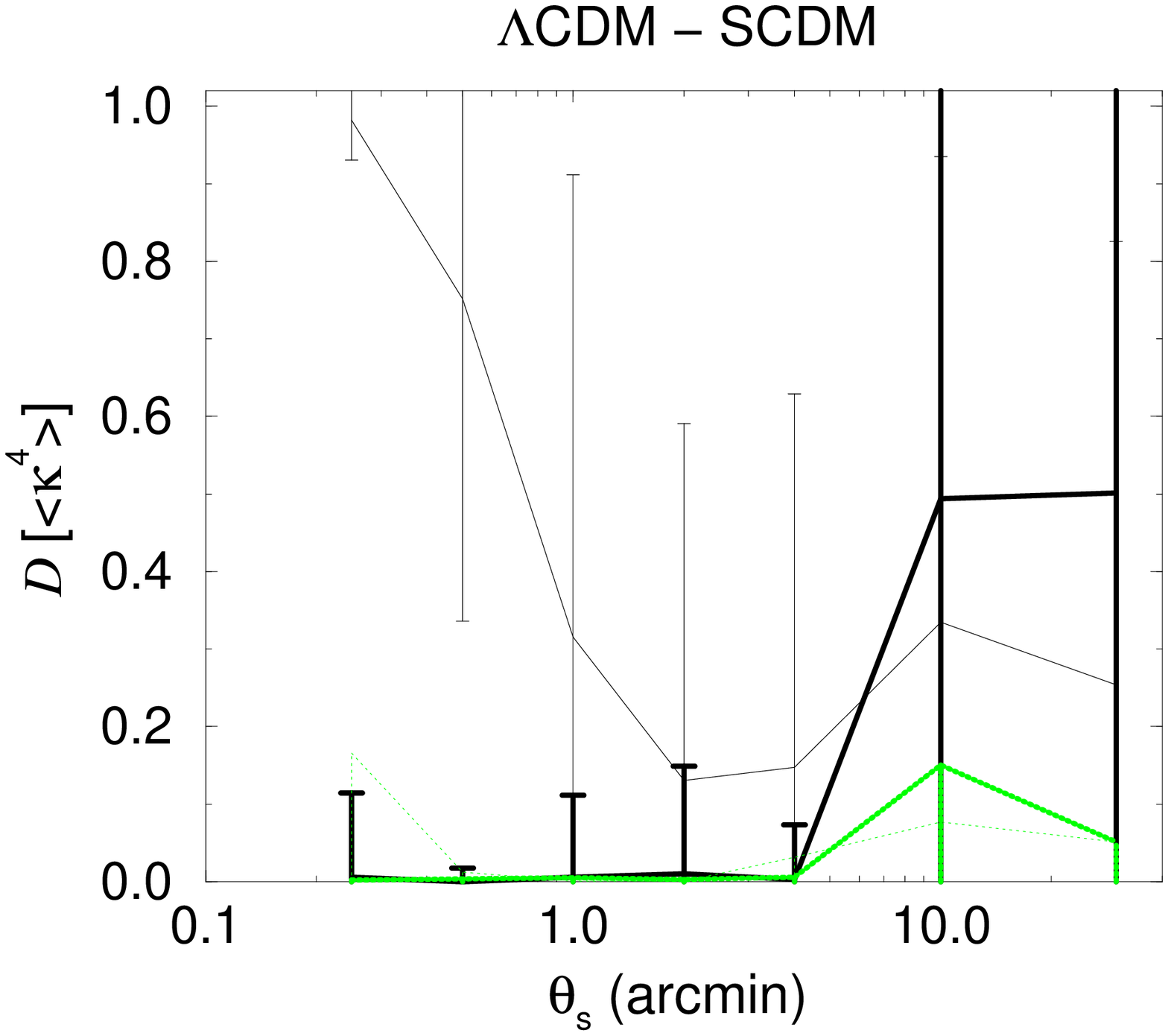}
\hspace{-1.2cm}
\epsfxsize=1.88in
\epsfbox{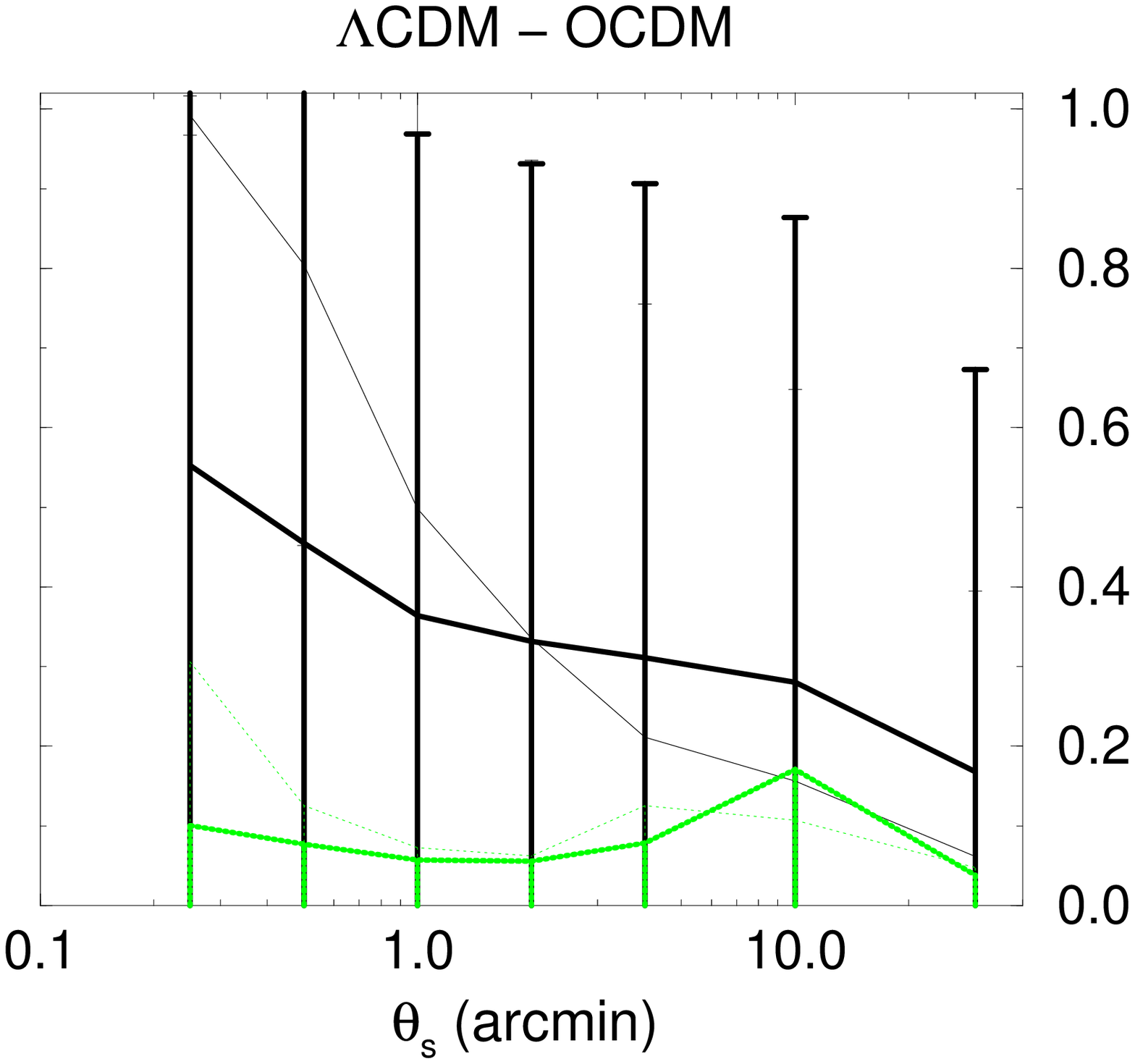}
\hspace{.5cm}
\epsfxsize=1.88in
\epsfbox{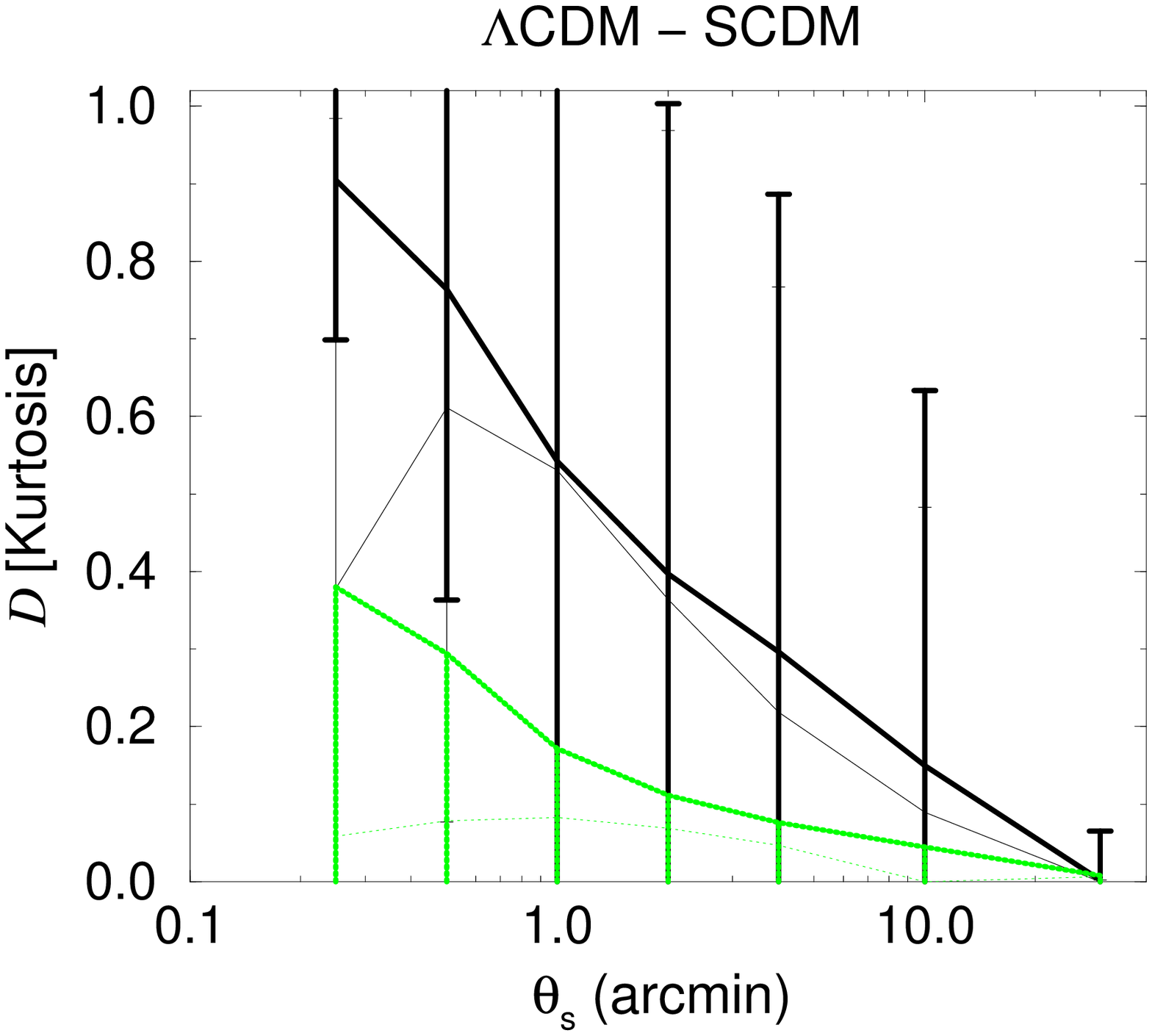}
\hspace{-1.2cm}
\epsfxsize=1.88in
\epsfbox{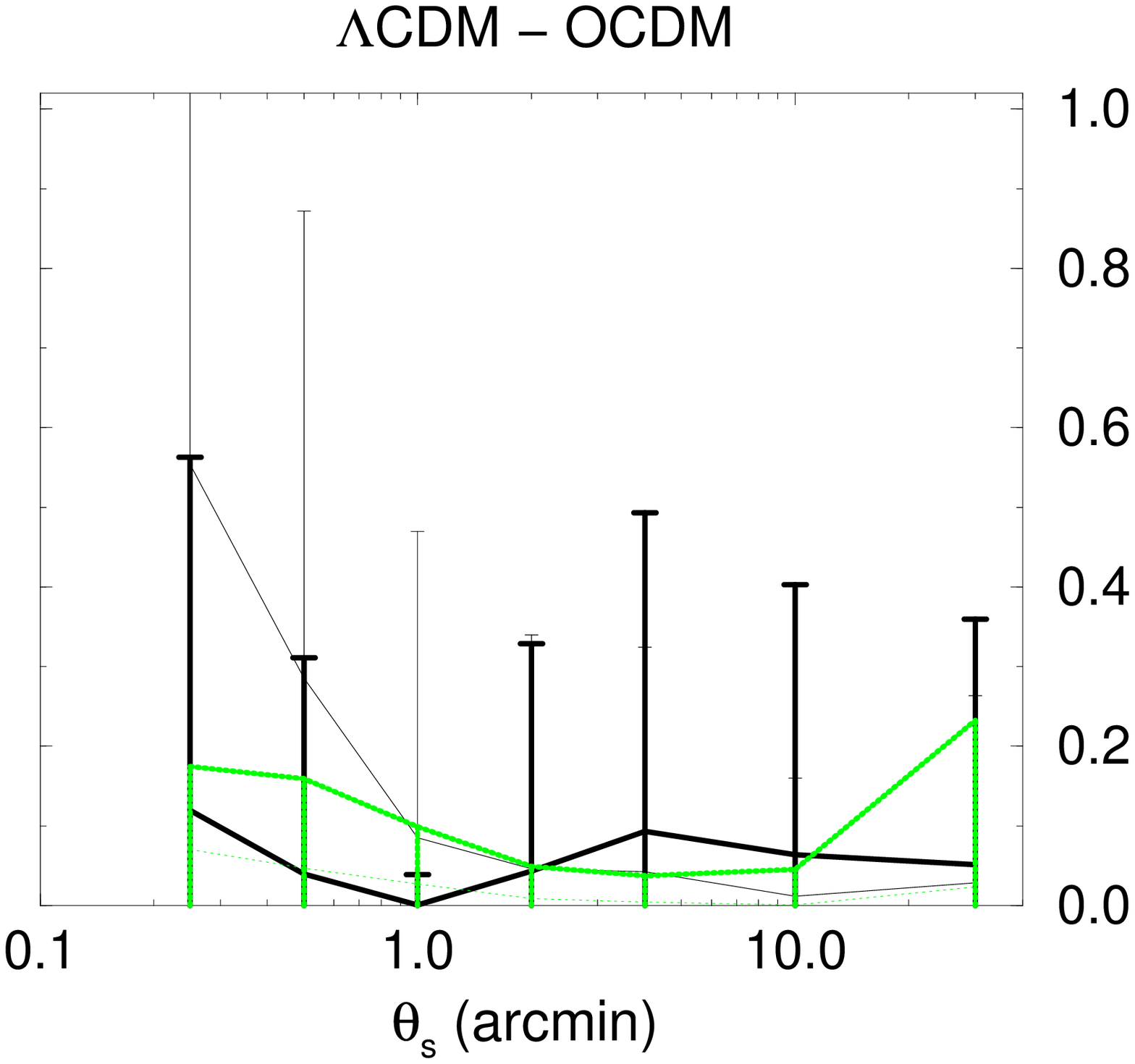}
\vspace{-0.2cm}
\caption{Convergence mean fourth power (left) and kurtosis (right). 
  Equal graphic conventions to figure \ref{k2-fig} are used.}
\label{k4-fig}
\end{figure}

The mean third power of the field is a direct measure of its
deviation from Gaussianity, and its value is not affected by the noise
field (see Figure \ref{k3-fig}), because $\skaco{n}= \skaco{ n^3}=0$,
\begin{equation}
\skaco{\kappa^3}= \skaco{\kappa^3_o} \; .
\end{equation} 
Unfortunately the noise field increases $\skaco{\kappa^3}$ error-bars,
reducing the differentiation between models
\begin{equation}
\Delta^2[\skaco{\kappa^3}] = \Delta^2[\skaco{\kappa_o^3}] + 
\Delta^2[\skaco{n^3}] + 2 \Delta^2[\skaco{\kappa_o^2 n}] +
2\Delta^2[\skaco{\kappa_o n^2}] \; .
\end{equation} 
Figure \ref{k3-fig} also shows the results for the related quantity
skewness, $S=\skaco{\kappa^3}/ \sigma^3_\kappa$. 
Its null differentiation between $\Lambda$CDM and OCDM, and not null
between $\Lambda$CDM and SCDM, suggests that the convergence skewness is
sensitive to the mass density, and insensitive to the cosmological
constant.

Figure \ref{k4-fig} shows the results for the mean fourth power of the
convergence field $\skaco{\kappa^4}$, and kurtosis 
$K=(\skaco{\kappa^4}/ \sigma^4_\kappa) - 3$. 
The contribution of the noise field can be visualized in the expression
\begin{equation}
\skaco{ \kappa^4} = \skaco{ \kappa^4_o } + 
\skaco{ n^4 } + 6 \skaco{ \kappa^2_o n^2} \; ,
\label{kappa4}
\end{equation}
where $\skaco{ n^4 }=3\sigma_n^4$. 
This coupling of the pure field with the noise field, 
$\skaco{ \kappa^2_o n^2}$ term,
allows the model differentiation for 
$\skaco{ \kappa^4}$ to be larger for the noisy maps than for the
pure ones. 
The kurtosis of Gaussian fields is null, and that is what is observed
for noisy convergence maps at small and large smoothing angles. 
At small $\theta_s$ because noise dominates, and at large $\theta_s$ because
of the extreme field smoothing.

\begin{figure}
\leavevmode
\epsfxsize=3.3in
\epsfbox{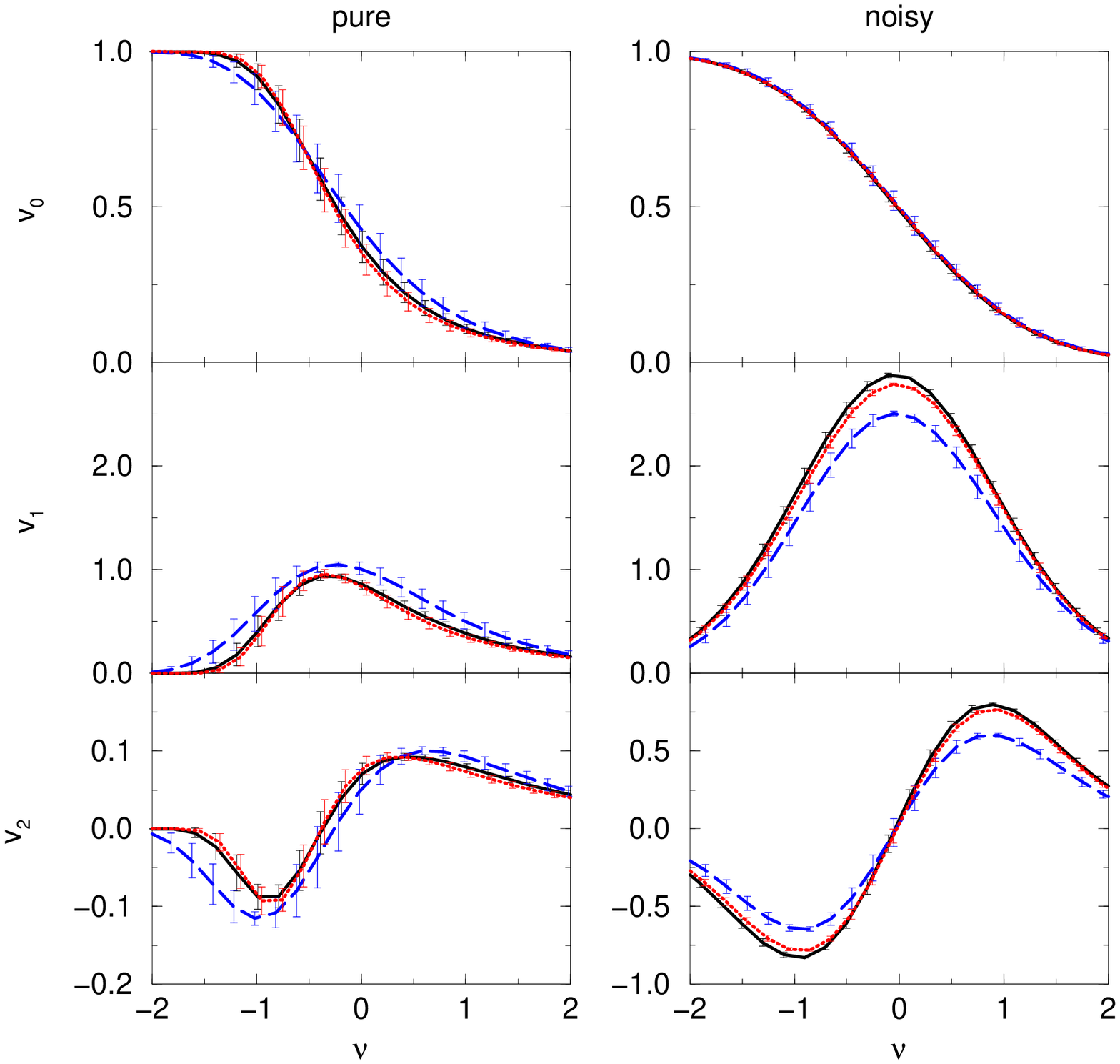}
\hspace{.3cm}
\epsfxsize=3.3in
\epsfbox{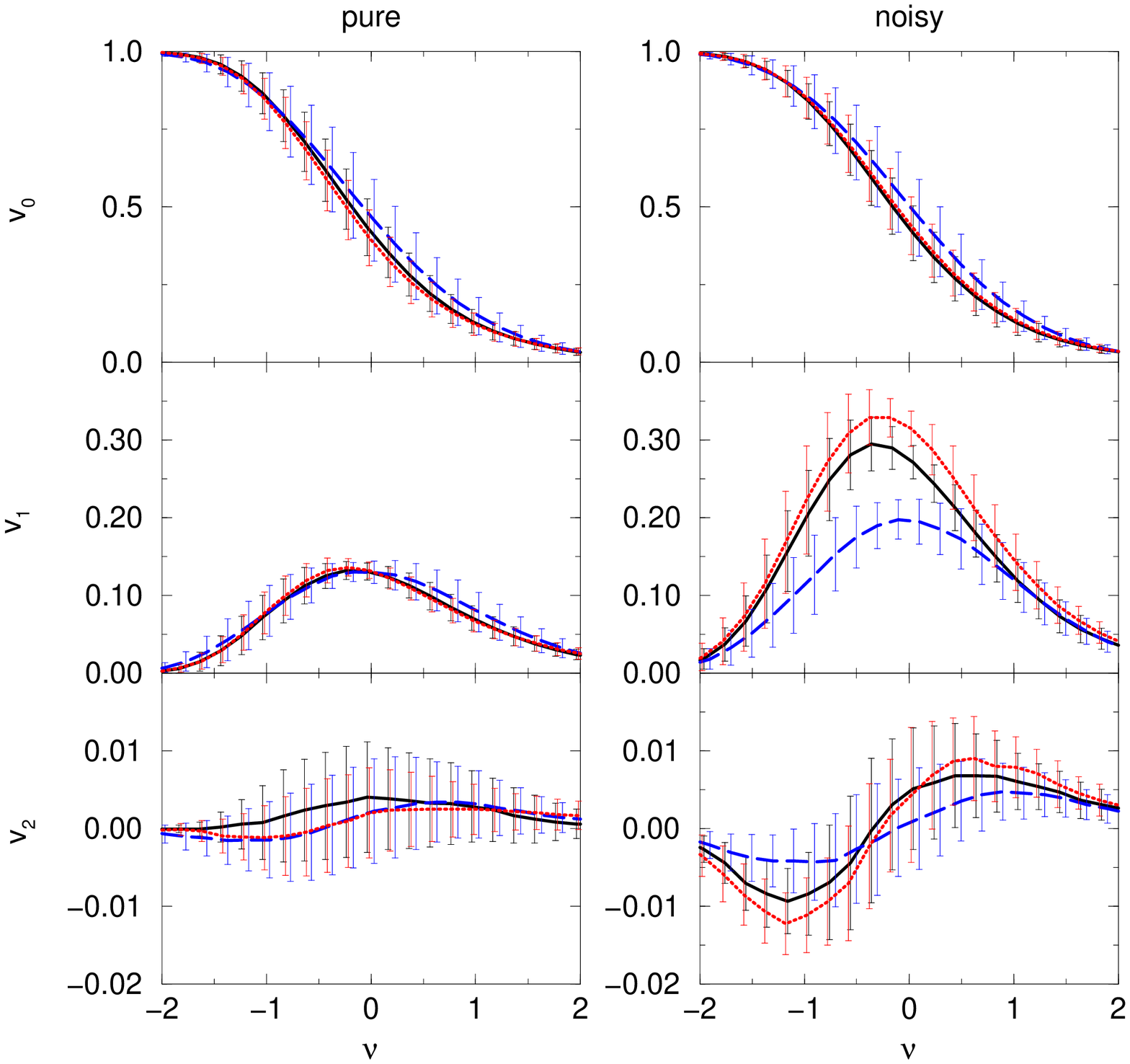}
\vspace{-0.4cm}
\caption{Minkowski functionals for the
  convergence field (pure and noisy), for smoothing angles 0.25 (left)
  and 4 (right) arcmin. 
  The first functional $v_0$ is dimensionless, $v_1$ units are
  arcmin$^{-1}$, and $v_2$ arcmin$^{-2}$.} 
\label{mink-fig}
\end{figure}
\begin{figure}
\leavevmode
\centerline{
\epsfxsize=1.8in
\epsfbox{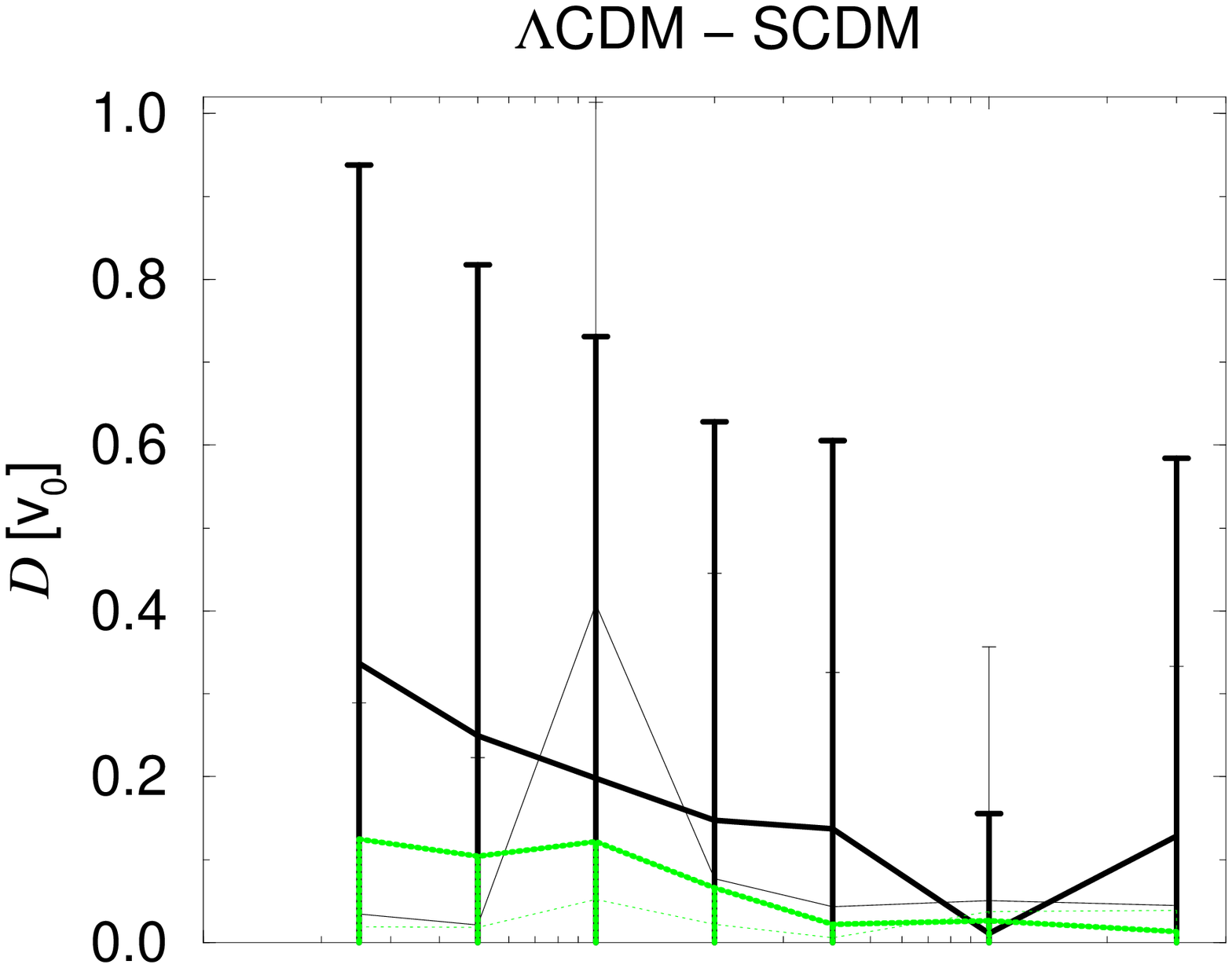}
\hspace{-1.2cm}
\epsfxsize=1.8in
\epsfbox{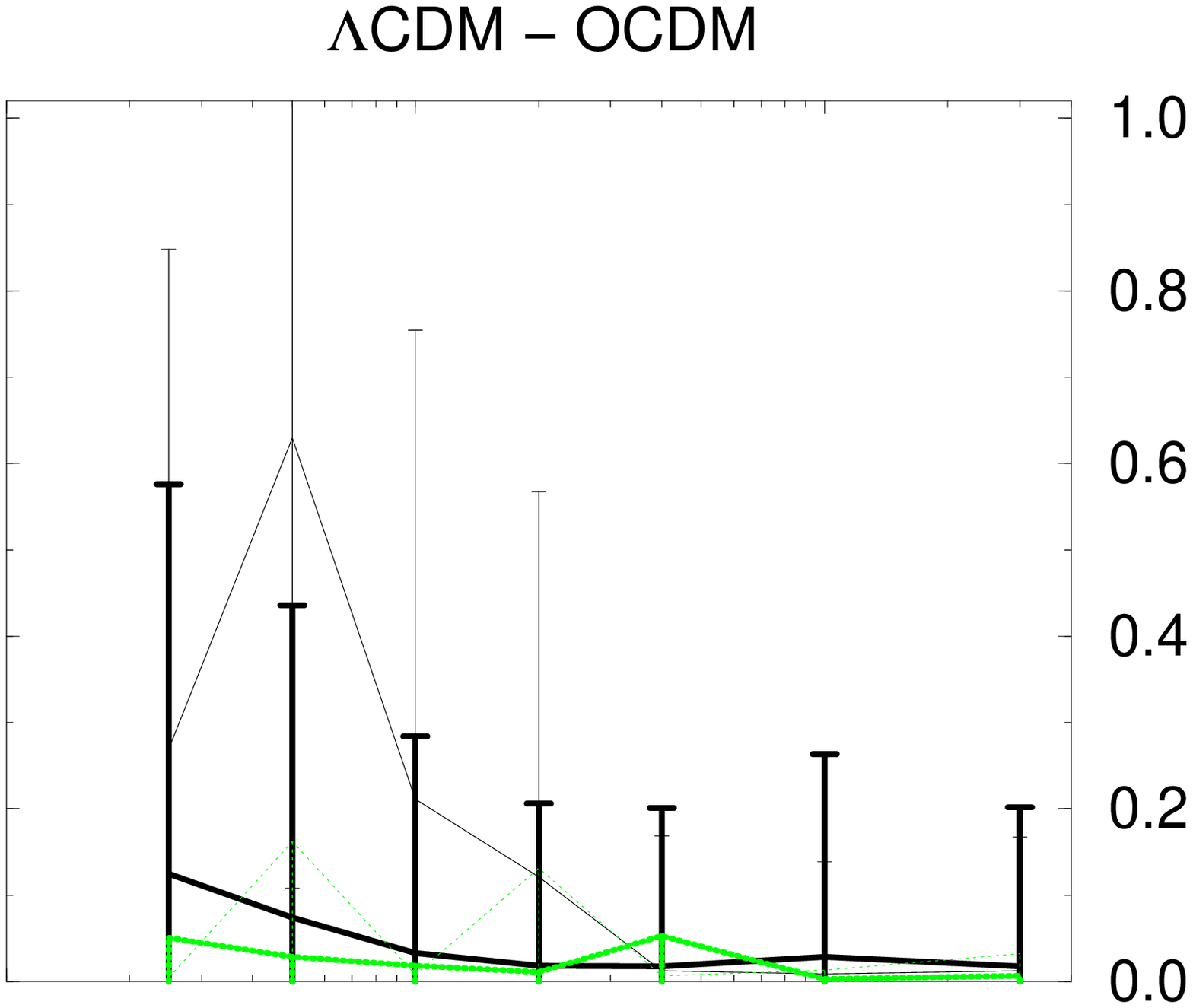}}
\vspace{-1.2cm}
\\
\centerline{
\epsfxsize=1.8in
\epsfbox{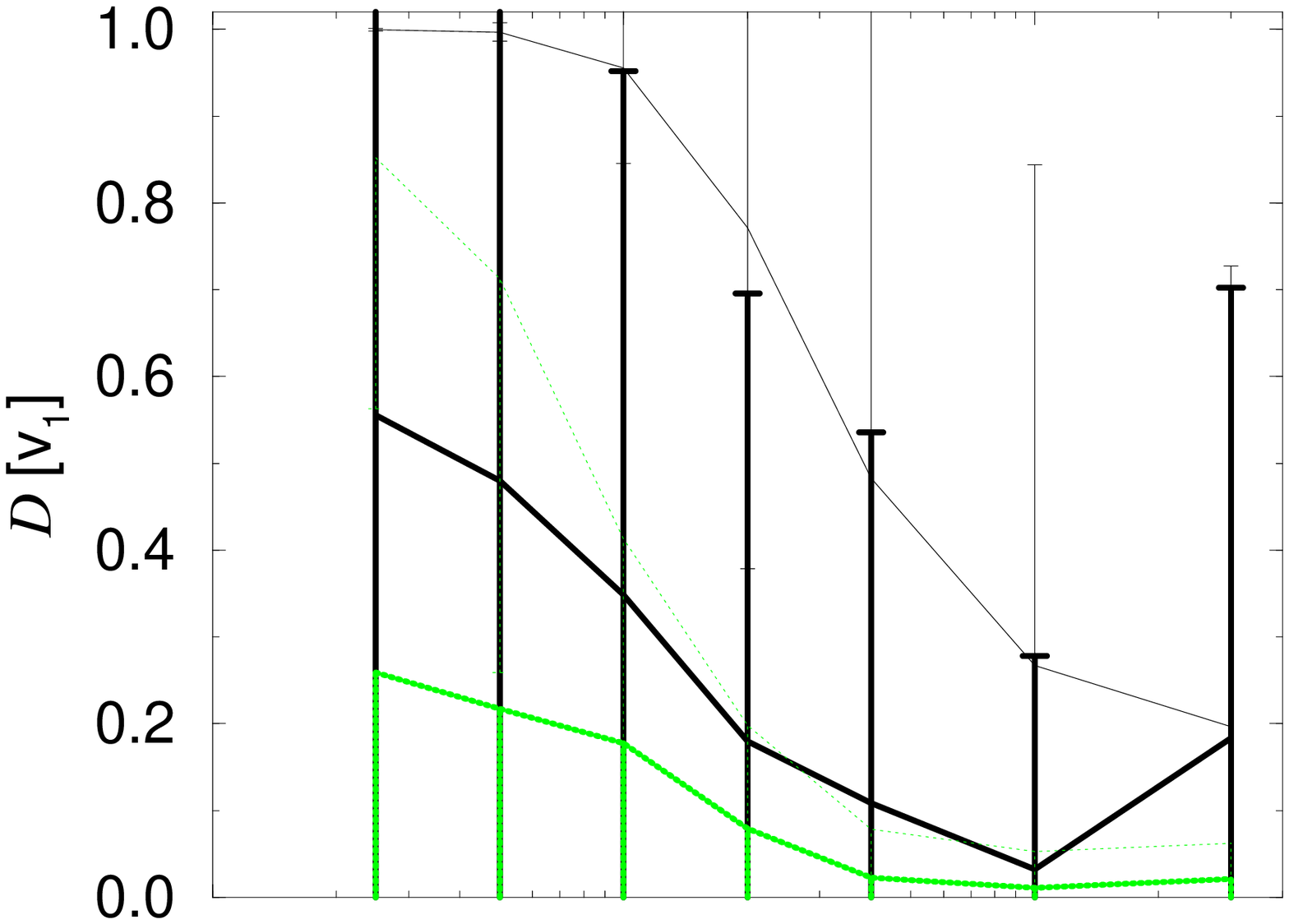}
\hspace{-1.2cm}
\epsfxsize=1.8in
\epsfbox{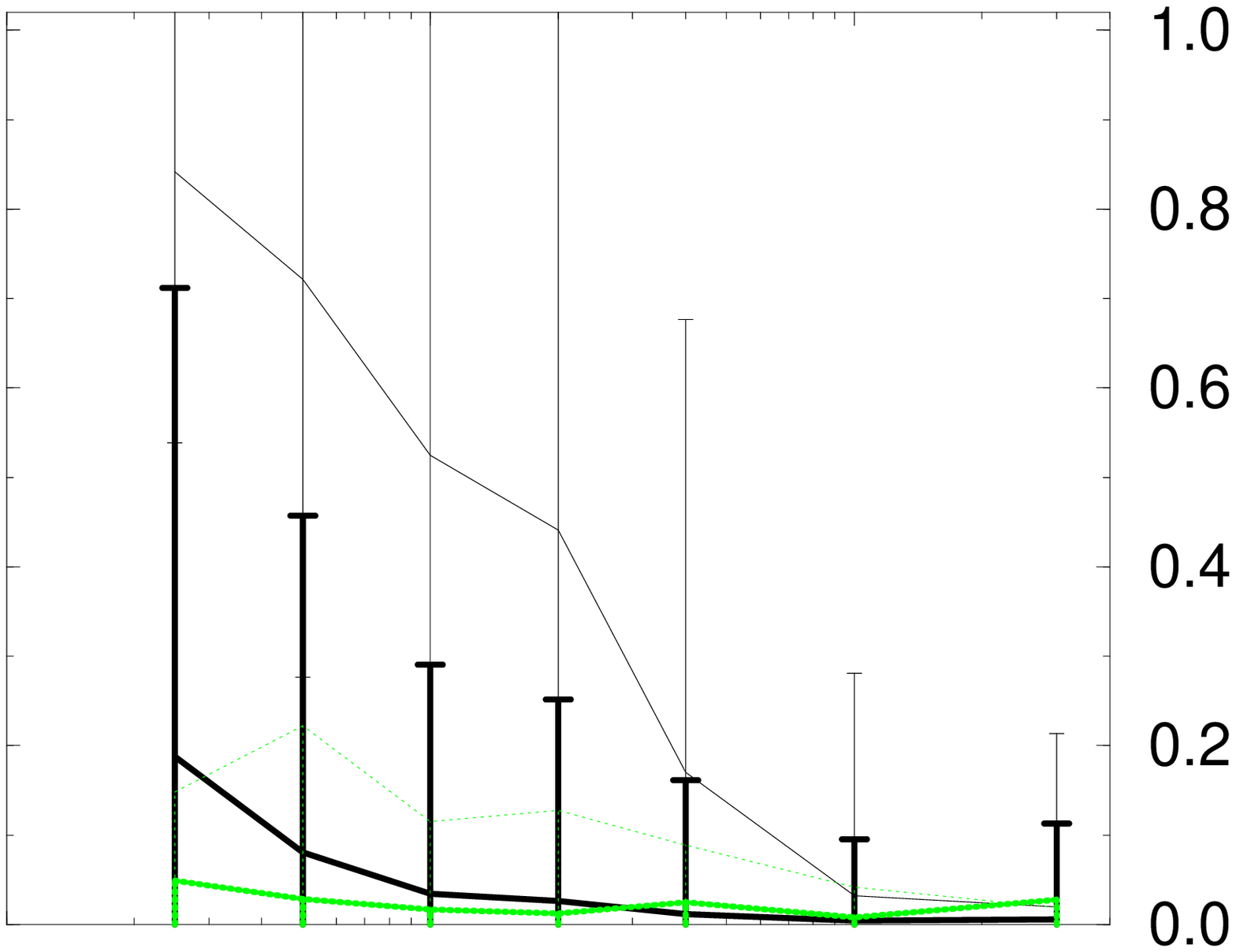}}
\vspace{-1.2cm}
\\
\centerline{
\epsfxsize=1.8in
\epsfbox{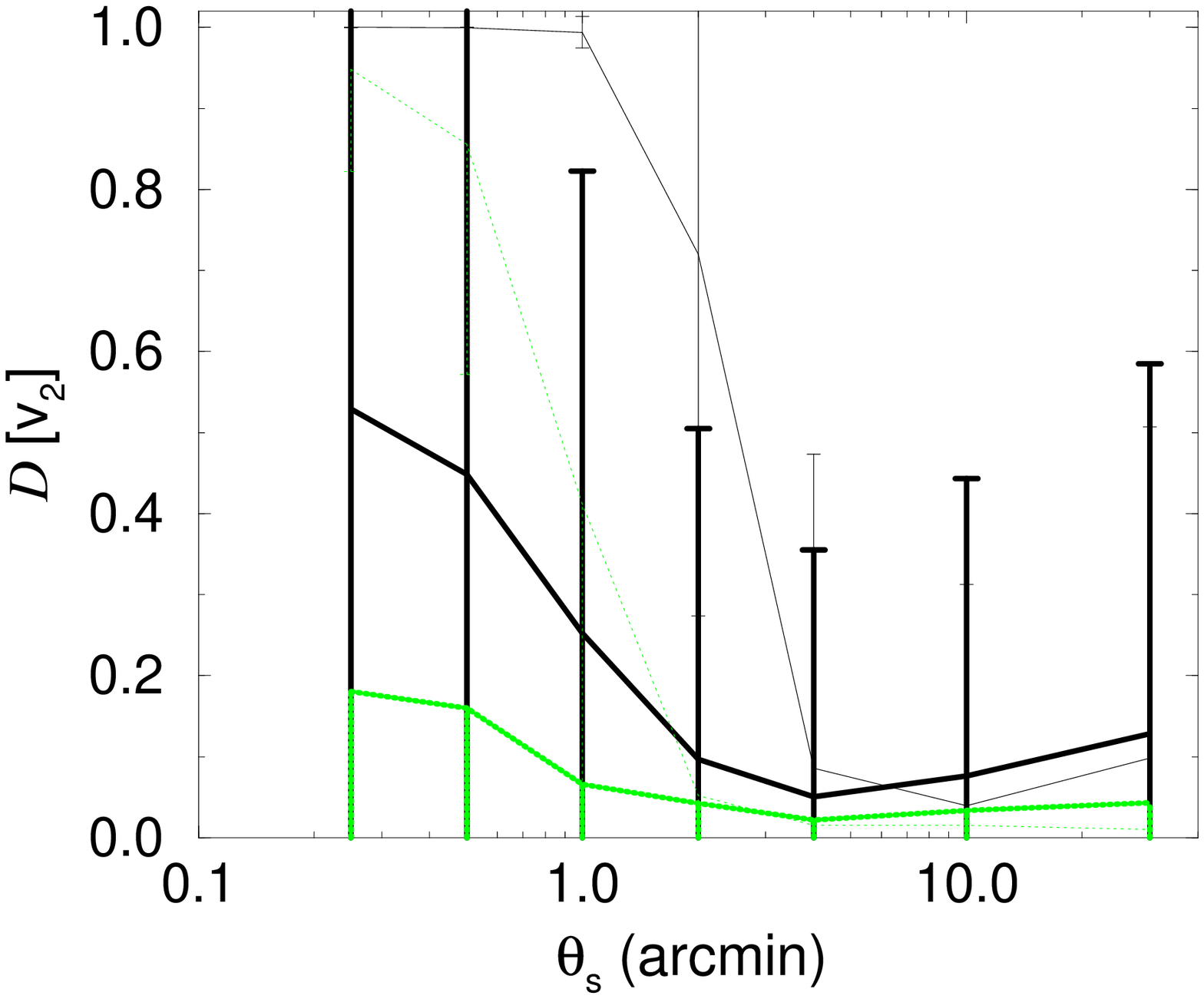}
\hspace{-1.2cm}
\epsfxsize=1.8in
\epsfbox{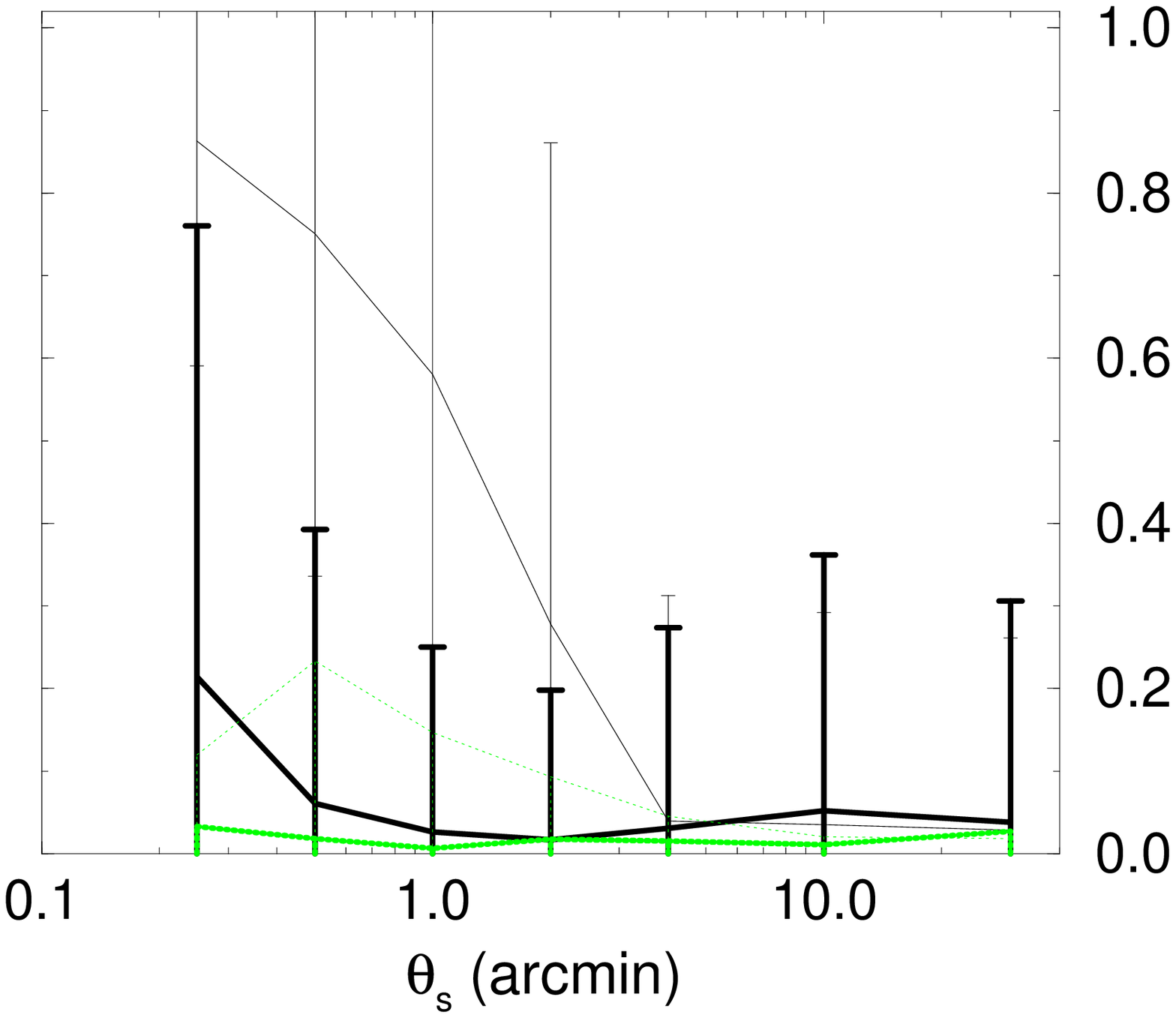}}
\vspace{-0.4cm}
\caption{Minkowski functional differentiation between
  $\Lambda$CDM-SCDM (left) and  
  $\Lambda$CDM-OCDM (right), as function of the smoothing
  angle $\theta_s$. Thick lines are for pure maps, and thin lines for
  noisy maps; solid lines are for 3$\times$3 degrees$^2$ fields, and
  dotted lines for 1$\times$1 degree$^2$ fields (the superior error-bars
  for these are not show for clarity).}
\label{mink-diff-fig}
\end{figure}

Some Minkowski functionals curves for the convergence field are shown
at Figure \ref{mink-fig}, and the differentiation results for this
analysis of the models are shown at Figure \ref{mink-diff-fig}.
Minkowski functionals are very sensitive to noise and smoothing. 
The morphology of the convergence maps for the examined models seem to
not differ greatly as described by the functionals. 
However for noisy fields the second and third functionals have
considerable discriminatory power at small smoothing angles.
It is in principle counter-intuitive that noisy maps would have a
larger differentiation than pure convergence maps, but as seen in the
case of the $\skaco{ \kappa^4}$ analysis, noise can introduce a
``statistical contaminant''. 
The discussion on Minkowski functionals of noisy maps presented in
Appendix \ref{mink-noisy} may be useful in understanding this effect.

Another interesting result to note is the low differentiation between
models under the first Minkowski functional analysis (fractional
area), which can be saw as a cumulative PDF. 
Figure \ref{PDF-fig} shows a strong differentiation between models
under the PDF analysis, so the results for ${\cal D}[v_0]$ could seem
contradictory with those for ${\cal D}[PDF]$, however when we remember
that Minkowski functionals are displayed as functions of the threshold
$\nu = \kappa / \sigma_\kappa$ this apparent discrepancy disappears. 
This observation suggests that the large PDF differentiation between
the models is due primarily to their very distinct convergence
variances.

\section{CONCLUSIONS}
\label{conclusions}

We study the potential of weak gravitational lensing maps to
differentiate between different cosmological models, using a variety
of statistical measures, including morphological analysis, and taking
in account the effects of noise and the uncertainties
resulting from a limited field size.

We reviewed how gravitational lensing maps can be generated, and also
revisited some statistics that can be used to characterize the lensing
maps, or extract information from them.  We introduced the quantity
$\cal D$, which quantifies the differentiation between two sets of
lensing maps under a given statistic.  We used simulations of
convergence maps in three cosmological models ($\Lambda$CDM, SCDM, and
OCDM) to study and compare these models.  Several of the analyses
considered here were also investigated in previous works, with results
consistent with ours.  The novelties presented in this paper are the
quantitative description of the differentiation between models (which
proved to be very useful), the possibility to compare side-by-side
different statistics, and a systematic consideration of the role of noise
and field size in all lensing measures.
It was also the first time that the Minkowski functionals for noisy
maps were calculated.

Our results show, as expected, that the lensing measures of small
fields have large variance, because of limited sampling.  These large
variances imply a small value for the differentiation $\cal D$: the
observation of a too small sky patch does not allow the discrimination
of cosmological models.  Other general behavior, independent of the
analysis or model considered, is the approximation of the
differentiation values obtained for pure and noisy convergence maps
when the smoothing angle is large.  As smoothing increases the noise
variance decreases  (noise becomes irrelevant for a sufficiently high
smoothing angle), but the pure convergence is also homogenized, and
the differentiation between models becomes small.  Our results suggest
that it is advantageous for model discrimination purposes to use a
minimum smoothing, even considering that this implies a noisier map.

For purposes of discriminating the three models examined in our
simulations, the variance and the angular power spectrum of the
convergence (two of the 
most simple and popular analysis) proved to be very good lensing
measures.  
Even for small field sizes, and noisy maps, the convergence
variance can differentiate between the models at a great confidence
level (if the smoothing is not too extreme).

Higher order statistics ($\skaco{\kappa^3}$, skewness, $\skaco{\kappa^4}$,
kurtosis, and Minkowski functionals) have higher variances (error
bars) in general, and lower discriminating power for the examined
models.  However, the information obtained from these statistics is
precious for its independence from the information offered by
statistics such as field variance, and angular power spectrum.  It is
possible to imagine two cosmological models that have the same mass
power spectrum, but different non-Gaussian features.  We found that
the second Minkowski functional ($v_1$) is an equal or better model
discriminant than the more commonly used third functional (the
topological genus), and is very competitive in relation to other
measures.
This suggests that Minkowski functionals should be included in the
row of available statistics to maximally extract information from
weak lensing maps. 
They should be particularly useful to differentiate models that have the
same PDF, but distinct morphology.
 
The presence of noise makes the extraction of information about
non-Gaussian features of the pure convergence much more intricate.
While for the analyses that are blind to non-Gaussianities (variance
and angular power spectrum) the noise term is clearly isolated from
the pure convergence term, for analyses sensitive to non-Gaussianities
(with exception of $\skaco{\kappa^3}$) the noise terms are entangled with
the pure convergence terms (see equations [\ref{pdf-convolution}],
[\ref{kappa4}], [\ref{noisy-v0}], and 
[\ref{noisy-S3}-\ref{noisy-Dk2}]).  
So, even for a simple noise model as the one used in our
simulations, the modification of the lensing measures due to noise is
nontrivial.  That implies that the retrieval of information  about the
pure convergence field (and ultimately, from the underlying
large-scale structure and cosmological model) from noisy maps is also
nontrivial - even more when the noise field is not well known.
Minkowski functionals are a particularly severe example of this
entanglement between the pure convergence with noise (as shown in Appendix
\ref{mink-noisy}). 

The complicated analytical description of lensing measures sensitive
to non-Gaussian features of noisy maps, 
and the intrinsic difficulty of incorporating 
observational aspects trough the analytical approach, 
favors the use of simulations over an
analytical approach for observational predictions, or parameter
estimation from real lensing maps. 
These arguments add to the work of Jain, Seljak \& White (2000), which
points to the limitations of perturbation theory in providing accurate
predictions for most weak lensing statistics at small scales (low
smoothing).

Different statistics of weak gravitational lensing maps reveal distinct
features of its originating cosmic mass distribution and
geometry. Therefore, a comprehensive and realistic study of the
underlying cosmological model through weak lensing requires the use of
a set of statistics, and the understanding of how these statistics and
their variances are affected by observational constraints such as field
size and the presence of noise.

\section*{ACKNOWLEDGMENTS}
I thank Robert Brandenberger for very helpful discussions, and
Uro\v{s} Seljak for providing codes and valuable knowhow. 
I also thank the Departments of Physics at Princeton and Rutgers for
the use of their facilities during the realization of this work.
The research at Brown was supported in part by the US Department of
Energy under Contract DE-FG0291ER40688, Task A.



\appendix

\section{Construction of the Differentiation}
\label{diff-constr}

\subsection{Simple case}

Let $A$ and $B$ be models or observations, $Y$ an analysis of lensing
maps, and $\sigma^2$ its variance. 
$Y_A$ and $Y_B$ are the result of applying the analysis $Y$ on
maps of $A$ and $B$. $Y_A$ is assumed to follow a
normal distribution of mean value $\bar{Y}_A$ and variance
$\sigma^2_A$
\begin{equation}
F[Y_A] = \frac{1}{\sqrt{2\pi}\sigma_A} e^{-(Y_A-\bar{Y}_A)^2/2\sigma_A^2} .
\end{equation}
Equivalently for $B$.

We define the analysis difference between the two models as
\begin{equation}
D = \frac{Y_A-Y_B}{\sqrt{\sigma_A^2+\sigma_B^2}} .
\end{equation}
It is easy to see that $D$ is a unit normal distribution,
\begin{equation}
F[D] = \frac{1}{\sqrt{2\pi}}
e^{-(D-\bar{D})^2/2},
\end{equation}
where the expectation value of $D$ is
\begin{equation}
\bar{D}= \frac{\bar{Y}_A-\bar{Y}_B}{\sqrt{\sigma_A^2+\sigma_B^2}} .
\end{equation}

In fact, we are interested in the absolute value of $D$, so the
distribution function for $|D|$, is
\begin{eqnarray}
F[|D|] &=& F[D] + F[-D] \\
&=& \frac{1}{\sqrt{2\pi}} 
\left[ e^{-(D-\bar{D})^2/2} + e^{-(D+\bar{D})^2/2}
\right] \; .
\nonumber
\end{eqnarray}

If $D=0$ we say the two models are similar. So the probability of $D=0$
gives a good measure of how similar the two models are under the
analysis $Y$:
\begin{equation}
F_{|D|}(0)= \sqrt{\frac{2}{\pi}} e^{-\bar{D}^2/2} ,
\end{equation}
which is normalized to 1 over integration in $\bar D$ (from 0 to $\infty$).

Two very different models ($\bar{D}\gg 0$) have
$F_{|D|}(0)\approx 0$; in contrast, two similar models have a large
probability that $|D|=0$. 
The maximum possible value for $F_{|D|}(0)$ is obtained when
$\bar{D}=0$, which  implies
\begin{equation}
\max \left[ F_{|D|}(0) \right] = \sqrt{\frac{2}{\pi}}
\end{equation}

We can finally define the {\bf differentiation} between the two models
under the analysis $Y$ as 
\begin{equation}
{\cal D}[Y] \equiv  1 - \frac{ F_{|D|}(0)}{\max [{F_{|D|}(0)}]}
= 1- e^{-\bar{D}^2/2} .
\label{differ}
\end{equation}

\subsection{Generalization}
We can generalize the differentiation for statistics that are
functions of a parameter $p$ (e.g. the angular power spectrum as
a function of the wavenumber $l$, or a Minkowski functional as
a function of the threshold $\nu$).
Note that $p$, and the analysis itself, can be a vector.

The same elements used before for the simple case
can be generalized to $Y_{A,B}(p)$, $\sigma_{A,B}(p)$,
\begin{equation}
D(p) = \frac{|Y_A(p)-Y_B(p)|}{[\sigma_A^2(p)+\sigma_B^2(p)]^{1/2}}. 
\end{equation}

Using a reasoning similar to the one used in the simple case, we can obtain
\begin{equation}
{\cal D}[Y] = 1 - e^{-\chi^2/2} ,
\label{differ-p}
\end{equation}
where
\begin{equation}
\chi^2 \equiv \left[ \int{W(p) \, dp} \right]^{-1} {\int{\bar{D}^2(p)
    W(p)  \, dp}} \; ,
\label{chisqgen}
\end{equation}
and $W(p)$ is a window function (which limits the parameter domain).

Note that the discrete form of $\chi^2$ assumes a very
recognizable representation (equation [\ref{chisquare}]). 
It becomes then clear that the
maximization of ${\cal D}[Y]$ is equivalent to the minimization of
$\chi^2$, best know as the least square method.

The differentiation can be further generalized to include the effects
of covariance by
\begin{equation}
\chi^2 = \frac{1}{N} \sum_{i,j}^N  
[\bar{Y}_A(p_i)-\bar{Y}_B(p_i)]
(\mbox{M}^{-1})_{ij}
[\bar{Y}_A(p_j)-\bar{Y}_B(p_j)] \, ,
\end{equation}
where
\begin{equation}
\mbox{M}_{ij} \equiv \skaco{[Y_A(p_i)-Y_B(p_i)]\,[Y_A(p_j)-Y_B(p_j)]}
\end{equation}
is the covariance matrix, and the brackets denote ensemble average. 
Note however that we did not include the effects of covariance in our
calculations in Section \ref{applex}.

\section{Minkowski functionals of noisy maps}
\label{mink-noisy}

The problem of expressing the Minkowski functionals for the noisy
convergence map $\kappa$ in terms of the functionals for the pure
convergence $\kappa_0$ and the smoothed noise $n$ seems to be an open
issue. 
For the first functional, the fractional area, we found a solution,
which can be obtained when one makes use of the fact that this
functional is the cumulative probability function
\begin{equation}
v_0(\nu)=\int_{\nu \sigma}^{+\infty}{F(x)dx} \; ,
\end{equation}
and writes the convergence probability distribution function
$F_\kappa$ as the convolution of $F_{\kappa_0}$ and $F_n$ (equation
[\ref{pdf-convolution}]), obtaining
\begin{equation}
v_0^{(\kappa)}(\nu) = - \int_{-\infty}^{+\infty}
{\left[ \frac{d}{d\nu^\prime}v_0^{(n)}(\nu^\prime) \right]
v_0^{(\kappa_0)} \left(\frac{\sigma_\kappa \nu - \sigma_n \nu^\prime
}{\sigma_{\kappa_0}} \right) d\nu^\prime } \,.
\label{noisy-v0}
\end{equation}

Another approach is to use the approximations
(\ref{mink0}-\ref{mink2}) and calculate how their terms change for a noisy
map. In fact, these approximations should be better representations
of the noisy map than of the pure map, because noise addition makes
the convergence closer to a Gaussian field.

The term $\sigma_1^2\equiv\skaco{(\nabla\kappa)^2}$ can be decomposed
as 
\begin{equation}
\skaco{(\nabla\kappa)^2} = 
\skaco{(\nabla\kappa_0)^2}+\skaco{(\nabla n)^2} \;,
\end{equation}
and it can be useful to note that 
\begin{equation}
\skaco{\nabla\kappa({\bm \theta}) \nabla\kappa({\bm \theta}^\prime) }
= - C_\kappa^{\prime\prime}(|{\bm \theta}^\prime-{\bm \theta}|) \; ,
\label{der-corr}
\end{equation}
where $C_\kappa^{\prime\prime}(r)$ is the second derivative of the
two-point correlation function. 
From (\ref{der-corr}) follows that 
$\skaco{(\nabla\kappa)^2}=- C_\kappa^{\prime\prime}(0)$.

The skewness parameters $S_3^{(0)}$ is altered by the addition of noise
by
\begin{equation}
S_3^{(0)}=\frac{\skaco{\kappa^3}}{(\sigma^2_{\kappa_0}+\sigma^2_n)^2}
\; ,
\label{noisy-S3}
\end{equation}
and to obtain $S_3^{(1)}$ and $S_3^{(2)}$ one needs to calculate 
\begin{equation}
\skaco{\kappa^2 (\nabla^2\kappa)} = 
\skaco{\kappa_0^2 (\nabla^2\kappa_0)} + \skaco{\kappa_0^2 (\nabla^2 n)}
+ \skaco{n^2 (\nabla^2\kappa_0+\nabla^2 n)} \,,
\end{equation}
\begin{equation}
\skaco{(\nabla \kappa \cdot \nabla \kappa)(\nabla^2\kappa)} =
\skaco{(\nabla \kappa_0 \cdot \nabla \kappa_0)(\nabla^2\kappa_0)} +
\skaco{(2 \nabla \kappa_0 \cdot \nabla n + \nabla n \cdot \nabla n)
(\nabla^2\kappa_0)} + 
\skaco{(\nabla \kappa +\nabla n) \cdot (\nabla \kappa +\nabla n)
(\nabla^2 n)} \; .
\label{noisy-Dk2}
\end{equation}

\end{document}